\documentclass[11pt]{article}

\usepackage{multirow}
\usepackage[nottoc]{tocbibind}

\usepackage{geometry}
\usepackage[affil-it, auth-sc]{authblk}
\usepackage[utf8]{inputenc}
\usepackage{graphicx}
\usepackage{amssymb}
\usepackage{amsthm}
\usepackage{amsmath}
\usepackage{enumitem}
\usepackage{color, soul}
\usepackage{verbatim}
\usepackage{lineno}
\usepackage{float}

\definecolor{darkred}{rgb}{0.9, 0.0, 0.0}
\definecolor{darkgreen}{rgb}{0.0, 0.5, 0.0}
\definecolor{darkred}{rgb}{0.5, 0.0, 0.0}
\usepackage[pdftex,colorlinks,bookmarks,linkcolor=darkred, citecolor=darkgreen]{hyperref}

\usepackage{bm}
\usepackage{feynmf}

\usepackage[sort&compress,numbers]{natbib}
\usepackage{eso-pic}
\usepackage{slashed}

\makeatletter
\newcommand{\vast}{\bBigg@{4}}
\newcommand{\Vast}{\bBigg@{5}}
\makeatother

\newcommand{\eq}[1]{\begin{equation} #1 \end{equation}}
\newcommand{\eql}[2]{\begin{equation} \label{eqn:#1} #2 \end{equation}}
\newcommand{\eqs}[1]{\begin{equation} \begin{split} #1 \end{split} \end{equation}}
\newcommand{\eqsl}[2]{\begin{equation} \label{eqn:#1} \begin{split} #2 \end{split} \end{equation}}

\newcommand{\eqnref}[1]{Eq.~(\ref{eqn:#1})}

\newcommand{\secref}[1]{Sec.~\ref{sec:#1}}

\newcommand{\appref}[1]{Appendix~\ref{app:#1}}
\newcommand{\figref}[1]{Fig.~\ref{fig:#1}}

\newcommand{\tabref}[1]{Table~\ref{tab:#1}}

\newcommand{\pr}[1]{\left(#1\right)}

\newcommand{\mean}[1]{\langle#1\rangle}

\newcommand{\ld}[1]{_{\mathrm{#1}}}
\newcommand{\mrm}[1]{\mathrm{#1}} 

\def\cut{\mathrm{cut}} 
\def\opt{\mathrm{opt}} 
\def\muH{\mu\mathrm{H}} 

\def\GEV{G_E^{V}} 
\def\GES{G_E^{S}}
\def\GMV{G_M^{V}} 
\def\GMS{G_M^{S}}

\newcommand\GeV{\, \mathrm{GeV}} 

\allowdisplaybreaks

\begin{document}

\AddToShipoutPictureFG*{

    \AtPageUpperLeft{\put(-60,-75){\makebox[\paperwidth][r]{FERMILAB-PUB-20-124-T}}}  
    }

\title{\bf Parameterization and applications of the low-$Q^2$ nucleon vector form factors}

\date{\today}

\author[1]{Kaushik Borah}
\affil[1]{Department of Physics and Astronomy, University of Kentucky, Lexington, KY 40506, USA}

\author[1,2]{Richard J.~Hill}
\affil[2]{Fermilab, Batavia, IL 60510, USA}

\author[3,4]{Gabriel Lee}
\affil[3]{Department of Physics, LEPP, Cornell University, Ithaca, NY 14853, USA}
\affil[4]{Department of Physics, Korea University, Seoul 136-713, Korea}

\author[1,2]{Oleksandr Tomalak}
\maketitle

\begin{abstract}
  We present the proton and neutron vector form factors in a
  convenient parametric form that is optimized for momentum transfers
  $\lesssim$ few GeV$^2$.  The form factors are determined from a
  global fit to electron scattering data and precise charge radius
  measurements.  A new treatment of radiative corrections is applied.
  This parametric representation of the form factors, uncertainties and correlations 
  provides an efficient means to evaluate many derived observables.
  We consider two classes of illustrative examples: neutrino--nucleon scattering cross sections at GeV energies for neutrino
  oscillation experiments and nucleon structure corrections for
  atomic spectroscopy.  
  The neutrino--nucleon charged current quasielastic (CCQE) cross section differs by 3\%--5\% compared to commonly used form factor models when the vector form factors are constrained by 
  recent high-statistics electron--proton scattering data from the A1 Collaboration.
    Nucleon structure parameter determinations include: the magnetic and Zemach radii of the proton and neutron, $[r_M^p, r_M^n] = [ 0.739(41)(23), 0.776(53)(28)]$~fm and $[r_Z^p, r_Z^n] = [ 1.0227(94)(51), -0.0445(14)(3)]$~fm; 
  the Friar radius of nucleons, $[(r^p_F)^3, (r^n_F)^3] = [ 2.246(58)(2), 0.0093(6)(1)]$~${\rm fm}^3$; 
  the electric curvatures, $[\langle r^4 \rangle^p_E, \langle r^4 \rangle^n_E ] = [1.08(28)(5), -0.33(24)(3)]$~${\rm fm}^4$; 
  and bounds on the magnetic curvatures, $[ \langle r^4 \rangle^p_M, \langle r^4 \rangle^n_M ] = [ -2.0(1.7)(0.8), -2.3(2.1)(1.1)]$~${\rm fm}^4$.
  The first and dominant uncertainty is propagated from the experimental data and radiative corrections, and the second error is due to the fitting procedure.
\end{abstract}

\newpage
\tableofcontents

\section{Introduction} 

A new generation of precision measurements, including
accelerator-based neutrino experiments and muonic atom
spectroscopy, demands a rigorous assessment of nucleon structure
parameters and their uncertainties.  The electromagnetic form factors
of the proton and neutron are critical inputs to searches for new
physical phenomena and to new precise measurements of the elementary
particles.  As one example, precise neutrino--nucleus interaction cross
sections are required in order to access fundamental neutrino
properties at long-baseline oscillation
experiments~\cite{Mosel:2016cwa,Katori:2016yel,Alvarez-Ruso:2017oui};
the electroweak vector form factors of the nucleons are an important
input to these cross sections, and are determined by an isospin
rotation of the electromagnetic form factors.  As another example,
muonic atom spectroscopy~\cite{pohl10, antognini13} has opened a new
window on the determination of fundamental constants, and has revealed
surprising discrepancies in comparisons of different approaches to
nucleon structure~\cite{Mohr:2015ccw}; it is critical to quantify
uncertainties of nucleon structure inputs for the muonic atom program
and also to incorporate constraints from these new measurements into
other processes, such as the above-mentioned neutrino cross sections. 

Recently, with Ye and Arrington~\cite{Ye:2017gyb}, two of us provided a new global fit of the proton
and neutron electromagnetic form factors, encompassing the entire
momentum transfer ($Q^2$) range of available elastic electron
scattering data.   That analysis provides a comprehensive discussion
of the available data, and the  fit provides a general purpose tool
for studying the form factors over a broad range of $Q^2$.  However,
the fit of Ref.~\cite{Ye:2017gyb} is not optimized for relatively
low-$Q^2$ applications, such as neutrino scattering in the GeV energy
regime.  First, the inclusion of very high-$Q^2$ data necessitates the 
introduction of a large number of parameters, many of which are 
irrelevant to low-$Q^2$
applications.  Second, the presentation of errors in
Ref.~\cite{Ye:2017gyb} (an envelope around the curve as a function of
$Q^2$)  does not allow a systematic propagation of errors into derived
observables.   Finally, since the focus of Ref.~\cite{Ye:2017gyb} was
in summarizing the implications of electron scattering data in
isolation, it did not incorporate the low-$Q^2$  constraint on the
proton electric form factor that emerges from muonic atom
spectroscopy.  While there is not a complete consensus regarding the
resolution of the so-called proton radius puzzle~\cite{pohl13,carlson15,Hill:2017wzi,Alarcon:2018zbz,Hammer:2019uab},
we believe it is important to be able to
consistently incorporate these data and study their impact 
for applications such as neutrino scattering. 

In this paper, we utilize the raw data selections and
uncertainty evaluations for electron scattering cross 
sections from Ref.~\cite{Ye:2017gyb} to present a
complete and compact parametric representation and covariance
matrix for the form factors suitable for GeV and sub-GeV scale applications.
Section~\ref{sec:FF} begins by describing 
the salient features of the data analysis and presenting the fit results. 
Section~\ref{sec:applications} considers several illustrative applications, beginning with a discussion of 
form factor radii and curvatures in Sec.~\ref{sec:curvature}.
Section~\ref{sec:ccqe} evaluates 
neutrino--nucleon scattering cross sections. 
Section~\ref{sec:atom} presents central values and uncertainties for several nucleon structure parameters that are
important for muonic atom spectroscopy.
Section~\ref{sec:summary} provides a summary discussion.
Appendix~\ref{app:syst} discusses tensions between datasets.
Appendix~\ref{app:TPE} provides details on the dispersive evaluation 
of two-photon exchange radiative corrections.
Appendix~\ref{app:complit} compares 
our numerical results for nucleon structure parameters 
to previous estimates. 

\section{Presentation of form factors} \label{sec:FF}

In this section, we begin by recalling definitions and conventions,
discuss our data selection and fit procedure, and present the fit results.

\subsection{Definitions}

The Dirac and Pauli form factors, $F^N_1$ and $F^N_2$, 
respectively, are defined as matrix elements of the electromagnetic current: 
\begin{align}\label{eq:Fi}
\langle N(p')|J_\mu^{\rm em}|N(p)\rangle=\bar u(p')
\left[\gamma_\mu \tilde{F}^N_1(Q^2)+\frac{i\sigma_{\mu\nu}}{2M_N} \tilde{F}^N_2(Q^2)q^\nu\right]u(p)\,,
\end{align}
where $q^\mu = p^{\prime \mu} - p^\mu$, 
$Q^2 = - q^2= -(p'-p)^2$, and $N$ stands for $p$ (proton) or $n$ (neutron). 
In the presence of radiative corrections, the on-shell form factors $\tilde{F}_i$ are
IR divergent, and we define IR finite
form factors $F_i(Q^2) \equiv F_i(Q^2,\mu=M_p)$ in the $\overline{\rm MS}$ scheme at the renormalization
scale $\mu=M_p$~\cite{Hill:2016gdf}.%
\footnote{
The IR finite form factors are defined in terms of a
standard factorization
formula: $\tilde{F}_i = F_{i,S}(Q^2,\mu) F_{i}(Q^2,\mu)$. 
Here $\tilde{F}_i$ is the (IR divergent) on-shell form factor;     
the soft function $F_{i,S}$ is IR divergent, but independent of hadron structure; 
and the hard function $F_{i}$ (also called a ``Born'' form factor in the literature)
is IR finite and encodes hadron structure.   In the following, $F_i(Q^2)$
refers to $F_i(Q^2,\mu = M_p)$. 
}
We will present results in terms of the Sachs electric and magnetic form factors,
which are related to the Dirac-Pauli basis by 
\begin{align}
G^N_E = F^N_1 - \frac{Q^2}{4M_N^2} F^N_2 \,, \quad
G^N_M = F^N_1 + F^N_2 \,. 
\end{align}
For some applications, it is convenient to work with
the isoscalar and isovector linear combinations, 
\begin{align}\label{isoG}
\GES = G_E^{p} + G_E^{n} \,, \quad
\GEV = G_E^{p} - G_E^{n} \,.  
\end{align}

The form factors can be expressed as a convergent expansion in the variable $z(Q^2)$ \cite{Hill:2010yb,Bhattacharya:2011ah,Lee:2015jqa},
\eql{zexp}{
G_E^N(Q^2) = \sum_{k=0}^{k_{\max}} a_k z(Q^2)^k \,,
\quad
G_M^N(Q^2) = G_M(0) \sum_{k=0}^{k_{\max}} b_k z(Q^2)^k \,,
\quad 
z(Q^2) = \frac{\sqrt{t_\cut +Q^2} - \sqrt{t_\cut - t_0}}{\sqrt{t_\cut +Q^2} + \sqrt{t_\cut - t_0}} \,.
}
The dimensionless coefficients $a_k$, $b_k$ in this expansion encode hadronic structure.
The parameter $t_\cut$ is the timelike kinematic threshold for particle production: 
$t_\cut = 9 m_\pi^2$ for isoscalar form factors and $t_\cut = 4 m_\pi^2$ for isovector form factors.%
\footnote{When the proton and neutron form factors are considered individually, the lower threshold $4m_\pi^2$ must be used.}
The parameter $t_0$ represents the point in the $Q^2$ plane mapping to $z=0$; 
this free parameter defines the expansion scheme and is chosen for convenience. 
For example, the choice of $t_0$ that ensures the smallest range of $|z|$ corresponding to $0<Q^2<Q^2_{\rm max}$ is
\eql{t0opt}{
t_0^\opt(t_\cut, Q^2_{\max}) = t_\cut (1 - \sqrt{1 + Q^2_{\max}/t_\cut}) \,.
}
Perturbative QCD requires that the form factors fall off faster than $1/Q^3$ in the large $Q^2$ limit~\cite{Lepage:1980fj},
which implies the four sum rules~\cite{Lee:2015jqa}
\eq{
\sum_{k=n}^\infty k(k-1) \cdots (k-n+1) a_k = 0 \,, \qquad n = 0,1,2,3 \,.   
}
It will also be useful to consider the small-$Q^2$ expansion of the form factors, written conventionally as 
\begin{eqnarray}
G^N_E(Q^2) &=& 
G^N_E(0) - \frac{\langle r^2 \rangle^N_E}{3!} Q^2+ \frac{\langle r^4 \rangle^N_E}{5!} Q^4 + \cdots \,, \label{Taylor_electric} \\
G^N_M(Q^2) &=& 
G^N_M(0) \Bigg ( 1 - \frac{\langle r^2 \rangle^N_M}{3!} Q^2+ \frac{\langle r^4 \rangle^N_M}{5!} Q^4 + \cdots \Bigg ) \,. \label{Taylor_magnetic}
\end{eqnarray}
We further define $r_E^N = \sqrt{\langle r^2 \rangle^N_E}$
and $r_M^N = \sqrt{\langle r^2 \rangle^N_M}$.\footnote{The notations $\langle r^2 \rangle^N_{E,M}$ and $\langle r^4 \rangle^N_{E,M}$ are motivated
  in a nonrelativisitic model with static charge distribution~\cite{Friar:1978wv}.
  We employ this common notation with the understanding that it is a purely conventional representation of the corresponding
  form factor derivatives, e.g., $\langle r^2 \rangle^N_E \equiv - 6\, dG_E^N/dQ^2|_{Q^2=0}$.}
  
\subsection{Data selection}

For elastic $ep$- and $en$-scattering measurements, 
a complete tabulation of the data and error assignments that we use 
can be found in the supplemental material of Ref.~\cite{Ye:2017gyb}.
We provide a short synopsis here.
The $ep$-scattering data are divided into three datasets: 
\begin{enumerate}[label=(\roman*)]
\item ``Mainz": the rebinned 2010 data  from the A1 experiment \cite{bernauer14} with modifications as detailed in Ref.~\cite{Lee:2015jqa}, which comprise 657 data points in the kinematic range $Q^2 < 1 \GeV^2$;
\item ``World'': the compilation of unpolarized cross-section data not contained in the ``Mainz" dataset~\cite{dudelzakphd, janssens66,
bartel66, albrecht66, frerejacque66, albrecht67, goitein67, litt70, goitein70, berger71, price71, ganichot72,
bartel73, kirk73, borkowski74, murphy74, borkowski75, stein75, simon80, simon81, bosted90, rock92, sill93,
walker94, andivahis94, dutta03, christy04, qattan05, bernauer14};
\item ``Pol'': $G_E^p/G_M^p$ ratios extracted from polarization data~\cite{milbrath99, pospischil01, gayou01, strauch02, punjabi05, maclachlan06,
jones06, crawford07, ron11, zhan11, paolone10, puckett12, puckett17}.
\end{enumerate}
For $en$ scattering, we include all data available from Refs.~\cite{meyerhoff94, eden94, passchier99, herberg99, rohe99,
golak01, schiavilla01, zhu01, bermuth03, madey03, warren04, glazier05, geis08, riordan10, schlimme13} for $G_E^n$ and
Refs.~\cite{rock82, lung93, gao94, anklin98, kubon02, anderson07, lachniet09} for $G_M^n$. 
As explained in detail in Appendix \ref{app:PRAD}, we do not include PRad data to the fit since complete uncertainty correlations are not yet available~\cite{Xiong:2019umf}.

In addition to electron scattering data, we include precision low-$Q^2$ constraints on the form factors. 
Charge conservation requires that $G_E^p(0)=1$ and $G_E^n(0)=0$. 
The magnetic moments of the proton and neutron determine~\cite{Mohr:2015ccw} (see also Ref.~\cite{Schneider:2017lff})
$G_M^p(0) \equiv  \mu_p/(e\hbar/2M_p) = 2.7928473508(85)$
and
$G_M^n(0) \equiv \mu_n/(e\hbar/2M_n) = -1.91304272(45)\times (M_n/M_p)$.%
\footnote{
  The $M_n/M_p$ factor results from a conventional expression of $\mu_n$ in
  units of the nuclear magneton, $e\hbar/2M_p$.  This difference is insignificant
  compared to other uncertainties in electron scattering fits,
  cf. footnote~4 of Ref.~\cite{Ye:2017gyb}.
  }
The proton electric charge radius can be inferred from the measurement of
the 2$S$--2$P$ Lamb shift in muonic hydrogen~\cite{pohl10};
we employ the updated value~\cite{antognini13}
\eql{rEpmuH}{
\pr{r_E^p}_{\muH} = 0.84087(39) \text{ fm} \,. 
}
The neutron electric charge radius is determined from neutron scattering length measurements 
on heavy targets~\cite{Kopecky:1995zz, Kopecky:1997rw},
which yield 
\eql{rEnscatt}{
  \langle r^2 \rangle^n_E = -0.1161(22) \text { fm}^2 \,.
}
We do not include external constraints on $r_M^p$ or $r_M^n$. We have not included dispersive constraints~\cite{Belushkin:2006qa,Lorenz:2014yda,Alarcon:2018zbz,Hammer:2019uab} on the form factors such as from $\pi\pi \to N\bar{N}$ data, since these constraints have either modest impact on the fits~\cite{Hill:2010yb} or introduce further theoretical considerations.  Our form factor results are presented with complete error budgets that may be compared to other determinations using dispersive analysis, lattice QCD or future electron scattering data.

\begin{table}[t!]
\centering
\begin{tabular}{|l|c|c|c|c|c|c|c|c|r|r|}
\hline
  Fit & $Q^2_{\max}$ [GeV$^2$] & Mainz & World & Pol & $G_E^n$ & $G_M^n$ & $r_E^p$ & $\langle r^2\rangle_E^n$
  & $\chi^2\quad$ & $n_{\rm dof}$ \\
\hline
$p$ & 1.0 & 657 & 0 & 0 & 0 & 0 & 1 & 0 & 475.35 & 650 \\
$n$ ($G_E^n$) & 1.0 & 0 & 0 & 0 & 29 & 0 & 0 & 1 & 14.81 & 26 \\ 
$n$ ($G_M^n$) & 1.0 & 0 & 0 & 0 & 0 & 15 & 0 & 0 & 8.03 & 11 \\ 
iso $(1\,{\rm GeV}^2)$ & 1.0 & 657 & 0 & 0 & 29 & 15 & 1 & 1 & 499.63 & 687 \\
iso $(3\,{\rm GeV}^2)$ & 3.0 & 657 & 480 & 58 & 37 & 23 & 1 & 1 & 1162.45 & 1241 \\
\hline
\end{tabular}
\caption{Number of data points from each dataset included in fits below the momentum transfer $Q^2_{\mathrm{max}}$.
The total $\chi^2$ and number of degrees of freedom of each fit are also displayed.
}
\label{tab:datapts}
\end{table}

We will consider two types of fits: first, a fit of separate proton and neutron data to their respective form factors; 
second, a fit of combined proton and neutron data to isospin-decomposed form factors.  
For our default proton fit (line 1 of Table~\ref{tab:datapts}), we employ the ``Mainz'' $ep$-scattering dataset in combination with the proton electric charge radius.
For our default neutron fit (lines 2 and 3 of Table~\ref{tab:datapts}), we consider $en$-scattering data with $Q^2 \leq 1\,{\rm GeV}^2$
in combination with the neutron electric charge radius.
For our default isospin-decomposed fit (line 4 of Table~\ref{tab:datapts}), we consider all of the above proton and neutron
data. 
Finally, we also consider an isospin-decomposed fit (line 5 of Table~\ref{tab:datapts})
that includes all of the above 
data, as well as neutron data with $1 \,{\rm GeV}^2 < Q^2 \leq 3\,{\rm GeV}^2$, 
and ``World'' and ``Pol'' data with $0 < Q^2 \leq 3\,{\rm GeV}^2$.
The total number of data points for each of these fits is summarized in Table~\ref{tab:datapts}. We also show the total $\chi^2$ and number of degrees of freedom for each fit.%
\footnote{The number of degrees of freedom, $n_{\rm dof}$, is equal to the sum of the number of data points from the respective row in the table, minus the number of form factor parameters (for definiteness, 
we count $k_{\rm max} -4$ parameters for each form factor that are not fixed by sum rules; cf. Table~\ref{tab:zexpchoices}).
Note that nuisance parameters in the data set (floating normalizations for all datasets except for $G_E^n$ and correlated systematic parameters for the Mainz dataset as described in Refs.~\cite{Lee:2015jqa,Ye:2017gyb}) are subject to $\chi^2$ constraints (i.e., each nuisance ``parameter" 
is accompanied by a corresponding ``data point"), and we do not include them in counting $n_{\rm dof}$.
}

\subsection{Radiative corrections to $ep$ scattering} \label{sec:radcorr}

One goal of the current work is a more robust accounting for 
radiative corrections to unpolarized $ep$ cross sections in the fits.%
\footnote{We follow the analysis of \cite{Ye:2017gyb,arrington07b, arrington07c}
  by omitting radiative corrections to the form factor ratios from polarization data, which are expected to be small compared to other uncertainties.}
Besides the standard QED corrections on the electron line, 
there are three types of radiative corrections that must be applied to 
scattering data in order to extract
the IR finite ``Born'' form factors defined after Eq.~(\ref{eq:Fi}).  
They may be classified as ``hadronic vertex'', ``hadronic vacuum polarization'' and ``two-photon exchange'' (TPE).  
The first, hadronic vertex, type of correction involves soft radiation and the shape of the event distribution as
a function of the inelasticity $\Delta E = E_e^{\prime \, \rm elastic} - E_e^\prime$ 
(where $E_e^\prime$ is the scattered electron energy and $E_e^{\prime \, \rm elastic}$ is the elastic limit). 
This correction is calculable from QED in the soft limit $\Delta E \ll m_\pi$, but is numerically
enhanced by large logarithms in that limit. 
In the Mainz data, the soft-photon tail was analyzed in detail but neglected higher-order
corrections that are larger than stated systematic uncertainties~\cite{Hill:2016gdf}; 
however, the bulk of these corrections is absorbed by floating normalization parameters. 
In the World data, uncorrelated uncertainties were included 
in the dataset~\cite{Ye:2017gyb} 
to account for possible model dependence in the treatment of the radiative tail.
In all cases, the error budgets from Ref.~\cite{Ye:2017gyb} are assumed to contain any residual
error from the approximate treatment of this correction; 
we include here the small discrepancy between the $\overline{\rm MS}$ 
convention defined after Eq.~(\ref{eq:Fi}),
and the commonly used Maximon--Tjon convention~\cite{Maximon:2000hm} for the soft 
subtraction (see Appendix~B of Ref.~\cite{Hill:2016gdf} for 
related discussion).
The second, hadronic vacuum polarization, type of correction was omitted from the Mainz dataset~\cite{bernauer14},
and treated nonuniformly in the World dataset.  
As with the hadronic vertex correction, the error budgets from Ref.~\cite{Ye:2017gyb} are assumed to contain any residual
error from approximate treatment of this correction
(see Sec.~1 of Ref.~\cite{Lee:2015jqa} for related
discussion).

The third, TPE type of correction, remains a significant contributor to the error budget for $ep$ scattering.
As with the hadronic vertex correction, the soft-photon part of the TPE correction is computable without
uncertainty in QED, while the remaining hard-photon part is removed according to
\begin{align}\label{eq:Born}
d\sigma^{\rm Born} &= \frac{d\sigma^{\rm expt}}{1 + \delta_{\rm TPE}}   \,. 
\end{align}
Here $d\sigma^{\rm expt}$ is the experimental cross section after
extracting leptonic QED corrections, and the above-mentioned 
hadronic vertex, vacuum polarization, and soft-photon TPE effects. The 
resultant $d\sigma^{\rm Born}$ is identified with
the tree-level (Mott) cross section computed using
Born form factors. 

At arbitrary $Q^2$, we account for differences between the true TPE correction and the previous default model employed in Ref.~\cite{Ye:2017gyb}, called there ``SIFF Blunden''~\cite{blunden05a}, by writing (for notational simplicity, 
we henceforth suppress the subscript ``TPE" on $\delta$)
\eqsl{TPE}{
  \delta &= \begin{cases}
    \delta^{\rm default} + x\, ( \delta^{\rm dispersive} - \delta^{\rm default} ) & Q^2 < 1 \GeV^2 \\
    \delta^{\rm default} + y\,  \delta^{\rm AMT} & Q^2 > 1 \GeV^2 \end{cases} \,.
}
For data below $Q^2=1\,{\rm GeV}^2$, we 
consider the dispersive analysis from
Refs.~\cite{Borisyuk:2008es,Tomalak:2014sva,Tomalak:2016vbf,Blunden:2017nby,Tomalak:2017shs,Tomalak:2018ere},
which determines $\delta = \delta^{\rm dispersive}$; 
details are provided in \appref{TPE}. 
We take the discrepancy between the default model and the dispersive analysis 
as an uncertainty and allow $x = 1 \pm 1$.
Above $Q^2=1$~GeV${}^2$, we consider the phenomenological correction from Ref.~\cite{arrington07c},
$\delta = \delta^{\rm default} + y \delta^{\rm AMT}$, 
which is designed to improve agreement between polarization measurements and 
TPE-corrected unpolarized Rosenbluth measurements at high $Q^2$;
the explicit form of $\delta^{\rm AMT}$ is provided in Appendix~\ref{app:TPE}.  
We take the discrepancy between the default model and this phenomenological
ansatz as an uncertainty and conservatively allow
$y = 1 \pm 2$.
Since the ansatz involving $\delta^{\rm AMT}$ is purely phenomenological, we  perform fits with $y=1\pm 2$ enforced as an uncorrelated error, as in Ref.~\cite{Ye:2017gyb}, whereas $x=1\pm 1$ is enforced as a correlated error. 
We have verified that taking $x$ uncorrelated or $y$ correlated does not significantly alter 
the results.%
\footnote{The results for the default proton and iso ($1 \GeV^2$) fits are $x=1.41(52)$ and $x=1.43(52)$, respectively.
For the iso ($3 \GeV^2$) fit treating $y$ as a correlated error, we obtain $x=2.17(41)$ and $y=1.74(60)$. 
Loosening the Gaussian bounds on $x$ by a factor of 5, the results for the default proton and iso ($1 \GeV^2$) yield small changes to $x=1.55(61)$ and $x=1.58(60)$, respectively.
If we do the same for $x$ and $y$ for the iso ($3 \GeV^2$) fit, we obtain $x=2.40(45)$ and $y=1.81(64)$.
}
As a practical summary, our treatment of radiative corrections follows 
Ref.~\cite{Ye:2017gyb} above $Q^2=1\,{\rm GeV}^2$, with the
additional parameter $x$ to describe TPE corrections below 
$Q^2 =1\,{\rm GeV}^2$.

\subsection{Fit parameters and procedure}

\begin{table}[t!]
\centering
\begin{tabular}{|c|c|c|c|c|}
\hline
form factor & $t_\cut$ [GeV$^2$] & $t_0$ [GeV$^2$] & $k_{\max}-4$ & $|a_k|_{\max}$ \\
\hline
$G_E^p, G_M^p$ & $4m_\pi^2$ & $-0.21$ & $4$ & 5 \\
$G_E^n, G_M^n$ & $4m_\pi^2$ & $-0.21$ & $4$ & 5\\
$\GES, \GMS$ & $9m_\pi^2$ & $-0.28$ & $4$ & 5 \\
$\GEV, \GMV$ & $4m_\pi^2$ & $-0.21$ & $4$ & 5 \\
\hline
\end{tabular}
\caption{Parameter choices for the $z$ expansion of the form factors in this paper.
  Throughout the paper, we use the charged pion mass $m_\pi = 0.13957$~GeV for the evaluation of $t_{\rm cut}$. 
  The values for $t_0$ are obtained by rounding $t_0^\opt(1 \GeV^2; 4m_\pi^2) \approx -0.21$,
  and $t_0^\opt(1 \GeV^2; 9m_\pi^2) \approx -0.28$.
}
\label{tab:zexpchoices}
\end{table}

Having defined our datasets and treatment of radiative corrections, let us determine the relevant parameters for the $z$-expansion analysis. 
We use data with $Q^2 \leq 1$~GeV${}^2$ for our default fits, and choose 
$t_0$ as in \eqnref{t0opt} to minimize
the maximum size of $|z|$ in this $Q^2$ range.
We enforce sum rules on the coefficients, fix the normalization of form factors at zero momentum transfer, and choose the number of 
free parameters in the $z$ expansion, $n_{\rm max} = k_{\rm max} -4 = 4$, sufficiently
large so that terms of order $|z|^{n_{\rm max} + 1}$ are small compared to experimental precision.%
\footnote{For the isovector threshold $t_{\rm cut}=4m_\pi^2$ and choice of $t_0=-0.21\,{\rm GeV}^2$, we have $|z|^5 < 0.0033$ when $0<Q^2<1\,{\rm GeV}^2$, and
$|z|^5 < 0.042$ when $0<Q^2<3\,{\rm GeV}^2$. 
For the isoscalar threshold  $t_{\rm cut}=9m_\pi^2$ and choice of $t_0=-0.28\,{\rm GeV}^2$, we have $|z|^5 < 0.0007$ when $0<Q^2<1\,{\rm GeV}^2$, and
$|z|^5 < 0.019$ when $0<Q^2<3\,{\rm GeV}^2$. 
}
Our results do not change significantly when $k_{\rm max}$ is increased; we illustrate this
in Sec.~\ref{sec:applications} by recomputing observables using $k_{\rm max} \to k_{\rm max}+1$.  
For all form factors, we have enforced Gaussian bounds,  $|a_k| \le 5$, $|b_k| \le 5$ ($k=1,\dots, k_{\rm max}$) on the coefficients (i.e., a term $a_k^2/5^2$ is included in the $\chi^2$ function).  
Our results do not change significantly when this bound is increased by a factor of 2.
In \tabref{zexpchoices}, we summarize the choices of $z$-expansion parameters used in our fits.
For each choice of dataset in Table~\ref{tab:datapts}, the fit returns form factors expressed as central values, errors, and correlations for the indicated number of 
free parameters.  

\subsection{Fit results} \label{sec:fitresults}

For the proton fit (line 1 of Table~\ref{tab:datapts}), 
the form factor coefficients are
\eqsl{coefGp}{
[a^p_1, a^p_2, a^p_3, a^p_4] &= [-1.4860(97), -0.096(52), 1.82(15), 1.29(41)]\,, \\
[b^p_1, b^p_2, b^p_3, b^p_4] &= [-1.464(11), 0.063(60), 1.74(21), -0.35(38)] \,.
}
For the neutron fits (lines 2 and 3 of Table~\ref{tab:datapts}), 
the form factor coefficients are
\eqsl{coefGn}{
[a^n_1, a^n_2, a^n_3, a^n_4] &= [0.084(18), -0.279(63), -0.15(32), 0.35(56)] \,, \\
[b^n_1, b^n_2, b^n_3, b^n_4] &= [-1.415(39), 0.22(17), 1.39(39), 0.0(1.5)]\,.
}
For the isospin-decomposed fit (line 4 of Table~\ref{tab:datapts}), 
the isovector form factor coefficients are
\eqsl{coefGV}{
[a^V_1, a^V_2, a^V_3, a^V_4] &= [-1.576(15), 0.177(77), 2.05(24), 0.88(57) ] \,, \\
[b^V_1, b^V_2, b^V_3, b^V_4] &= [-1.456(13), 0.186(67), 1.63(23), -0.73(46)] \,,
}
and the isoscalar form factors are given by 
\eqsl{coefGS}{
[a^S_1, a^S_2, a^S_3, a^S_4] &= [-1.809(17), 0.91(12), 1.92(27), -0.98(82) ] \,, \\
[b^S_1, b^S_2, b^S_3, b^S_4] &= [-1.938(57), 0.78(25), 3.71(88), -4.0(2.8)] \,.
}

Whereas in Ref.~\cite{Ye:2017gyb} we considered the most inclusive dataset, 
here we have chosen the default proton dataset to contain the most recent precise measurements and to minimize internal data tensions.  
For definiteness we have included neutron data up to the same $Q^2_{\rm max}=1\,{\rm GeV}^2$. 
We remark that the Mainz dataset predicts 
$r_E^p=0.879(18)\,{\rm fm}$ when the $\mu$H charge radius constraint is removed~\cite{Lee:2015jqa}; 
this value is in only mild tension, $2.2\,\sigma$, with \eqnref{rEpmuH}.%
\footnote{This current fit corresponds to the ``alternate approach" described in 
Sec.~VI.C.3 of Ref.~\cite{Lee:2015jqa}, which yielded $r_E^p=0.891(18)\,{\rm fm}$
(line 7 of Table~XIV).  The small difference with 0.879(17) results from omitting 
sum rule constraints on the coefficients, omitting the floating TPE correction 
in \eqnref{TPE}, and restricting to $Q^2_{\rm max}=0.5\,{\rm GeV}^2$.}
The absence of more severe internal data tensions does not guarantee the
absence of potentially underestimated systematics; for a fuller discussion 
we refer to Ref.~\cite{Lee:2015jqa}.

Plots in Appendix~\ref{app:syst} compare our $G_E^p, G_M^p$, $G_E^n, G_M^n$, and $G_E^V, G_M^V$ form factors
against those of our previous global fit in Ref.~\cite{Ye:2017gyb}
and to the BBBA2005 parameterization of Ref.~\cite{bradford06}. 
In Supplemental Material, we provide values for the coefficients and covariance matrices suitable for precise evaluation of charge radii and other physical quantities,%
\footnote{The linear combination of coefficients that defines the charge radius is more precisely determined with the form factor parameters and the covariance matrix from the Supplemental Material than by evaluation using \eqnref{coefGp} and neglecting correlations.}%
as well as values for the coefficients from a fit with $k_\mathrm{max} = 9$ which we use in applications to estimate the error from $z$-expansion truncation.

\section{Illustrative applications} \label{sec:applications}

Having determined the form factor coefficients, errors and correlations, 
let us illustrate with some relevant physical examples. 
We begin in Sec.~\ref{sec:curvature} by evaluating form factor radii and curvatures.  Section~\ref{sec:ccqe} discusses neutrino scattering applications and Sec.~\ref{sec:atom} considers nucleon structure parameters for atomic spectroscopy. 

\subsection{Form factor radii and curvatures} \label{sec:curvature}

\begin{table}[tb]
\centering
\caption{Electric and magnetic radii of proton and neutron using 
form factor parameters and bounds of Table~\ref{tab:zexpchoices} and 
datasets of Table~\ref{tab:datapts}.  For $r_M$, the second number in each table 
results from changing the default $k_{\rm max}=8$ to $k_{\rm max}=9$.
}
\vspace{3mm}
\label{radii0}
\begin{minipage}{\linewidth}  
\centering
\begin{tabular}{|l|c|c|c|c|c|c|c|c|c|}
  \hline          
  Fit choice  &  $r_E^p$ $[\mathrm{fm}]$ & $r^p_M$ $[\mathrm{fm}]$ & $\mean{r^2}^n_E$ $[\mathrm{fm^2}]$ & $r^n_M$ $[\mathrm{fm}]$
\\
\hline
$p$ & $0.84089(39) $ & $0.739(41) $, $0.716(44)$& --- & --- \\
$n\, (G_E^n)$ & --- & ---& $-0.1161(22)$ & ---\\
$n\, (G_M^n)$ & --- & --- & --- & $0.881(83)$, $0.878(79)$ \\
iso (1~GeV$^{2}$) &$0.84090(39) $&$0.749(36) $, $0.729(38) $&$-0.1160(22) $&$0.776(53) $, $0.748(57) $\\
iso (3~GeV$^{2}$) &$0.84097(39) $&$0.799(23) $, $0.819(25) $&$-0.1160(22) $&$0.821(34) $, $0.855(38) $\\
\hline 
\end{tabular}
\end{minipage}
\end{table}

\begin{table}[h]
\centering
\caption{Same as Table~\ref{radii0}, but for curvatures.}
\vspace{3mm}
\label{curvatures0}
\begin{minipage}{\linewidth}  
\centering
\begin{tabular}{|l|c|c|c|c|c|c|c|c|c|}
  \hline          
  Fit & $\mean{r^4}^p_E$ $[\mathrm{fm^4}]$ & $\mean{r^4}^p_M$ $[\mathrm{fm^4}]$ & $\mean{r^4}^n_E$ $[\mathrm{fm^4}]$ & $\mean{r^4}^n_M$ $[\mathrm{fm^4}]$
\\
\hline
$p$ & $1.08(28) $, $1.13(30)$ & $-2.0(1.7) $, $-2.8(1.8)$& --- & --- \\
$n\, (G_E^n)$& --- & --- & $-0.37(62)$, $-0.35(63)$ & --- \\
$n\, (G_M^n)$ & --- & --- & --- & $1.6(3.3)$, $1.4(3.3)$ \\
iso (1~GeV$^{2}$) &$1.25(23) $, $1.21(23) $&$-1.6(1.5) $, $-2.4(1.5) $&$-0.33(24) $, $-0.30(25) $&$-2.3(2.1) $, $-3.4(2.2) $\\
iso (3~GeV$^{2}$) &$0.83(18) $, $0.78(19) $&$-0.6(9) $, $0.6(1.2) $&$0.04(20) $, $0.08(21) $&$-1.1(1.3) $, $0.8(1.7) $\\
\hline 
\end{tabular}
\end{minipage}
\end{table}

The nucleon radii, defined in Eqs.~(\ref{Taylor_electric}) and (\ref{Taylor_magnetic}),
are presented in Table~\ref{radii0}, where each line represents the result of the fit
using the corresponding dataset in Table~\ref{tab:datapts}. 
For each entry in the table, the first number represents the fit with 
default $k_{\rm max}=8$ (as in Table~\ref{tab:zexpchoices}), 
and the second number represents the fit with $k_{\rm max}= 9$. 

As expected, the output proton and neutron electric radii are driven by
the precise external constraints on these quantities and the $k_{\rm max}$ dependence is insignificant and not displayed.%
\footnote{Removing the neutron charge radius constraint from the PDG in \eqnref{rEnscatt}, the $n$~$(G_E^n)$ fit as in Table~\ref{tab:datapts} yields 
$\langle r^2 \rangle^n_E = -0.075(95)(5) \text{ fm}^2$, 
the iso ($1 \GeV^2$) fit yields $\langle r^2 \rangle^n_E = -0.092(34)(1) \text{ fm}^2$, 
and the iso ($3 \GeV^2$) fit yields $ \langle r^2 \rangle^n_E = -0.100(28)(8) \text{ fm}^2$.}
The determination of $r_E^p$ using low-$Q^2$ data released in 2019 by the PRad experiment 
is discussed in Appendix~\ref{app:syst}.  

The proton and neutron magnetic radii are consistent between fits, and represent 
the best values for these quantities obtained from electron scattering data plus external charge radius constraints.
The proton magnetic radius from the default fit, $r_M^p = 0.739(41)(23)\,{\rm fm}$,
should be compared to (and does not essentially alter) 
our previous extraction from 2010 Mainz data, 
$r_M^p=0.776(34)(17)\,{\rm fm}$.%
\footnote{
A naive average with the analogous fit to World data without Mainz data, 
$r_M^p=0.914(35)\,{\rm fm}$, is used to arrive at the PDG recommended value
$r_M^p=0.851(26)\,{\rm fm}$~\cite{Tanabashi:2018oca}.
Our current fit corresponds to the alternate approach described in Sec.~VI.C.3 of Ref.~\cite{Lee:2015jqa}, which yielded $r_M^p=0.792(49)\,{\rm fm}$
(line 7 of Table~XIV).}  
The difference results from omitting the muonic hydrogen constraint of \eqnref{rEpmuH}, omitting the floating TPE correction in \eqnref{TPE},
omitting sum rule constraints on the coefficients, 
and restricting to $Q^2_{\rm max}=0.5\,{\rm GeV}^2$.%
\footnote{
Removing the $\mu$H constraint from our default fit shifts the central value 
by $\sim 0.7\,\sigma$: $r_M^p = 0.739(41)\,{\rm fm} \to 0.768(42)\,{\rm fm}$; 
further removing the floating TPE parameter, the result would be $0.774(41)\,{\rm fm}$.
}
The neutron magnetic radius from our default fit, $r_M^n=0.776(53)(28)\,{\rm fm}$, 
represents a new extraction.
Our value may be compared to the result 
$0.89(3)\,{\rm fm}$ from Ref.~\cite{Epstein:2014zua}
that performed a $z$-expansion fit to $ep$ and $en$ scattering data 
and $\pi\pi \to N\bar{N}$ data,  
utilizing a dataset for $G_M^p$ from Ref.~\cite{arrington07c} 
that did not include 2010 Mainz data.  
The PDG recommended value $r_M^n=0.864(9)\,{\rm fm}$~\cite{Tanabashi:2018oca} 
results from a naive  
average of this result and the result $r_M^n=0.862(9)\,{\rm fm}$ from 
Ref.~\cite{Belushkin:2006qa} that performed a global fit of 
spacelike and timelike data to model spectral functions.

The form factor curvatures, $\mean{r^4}$ from Eqs.~(\ref{Taylor_electric}) and (\ref{Taylor_magnetic}),
are presented in Table~\ref{curvatures0}.
Only the proton electric curvature is determined to be nonzero with statistical significance.
As for the radii, different fit variations are consistent within uncertainties.  
We provide previous estimates of curvature in \tabref{curvatureslit} of Appendix~\ref{app:complit}.

For both radii and curvatures, the iso~(1~GeV$^2$) fit yields a modest reduction in uncertainty
compared to the separate proton and neutron fits, 
which can be traced to more data and a higher threshold $t_{\rm cut}$ 
(hence a smaller range of $|z|$ and a smaller number of relevant form factor coefficients) 
in the isoscalar channel.  
There is further reduction in the errors using the iso~(3~GeV$^2$) fit due to the inclusion of more data.
As discussed in Appendix~\ref{app:syst}, 
these additional data introduce a significant and unresolved 
tension with the Mainz dataset; 
we focus on the $p$, $n$ and iso~($1 \GeV^2$) fits as our default results.

\subsection{Neutrino--nucleon scattering \label{sec:ccqe}}

The elementary signal process for neutrino oscillation experiments is charged current quasielastic scattering,%
\footnote{For a classic review see Ref.~\cite{LlewellynSmith:1971uhs}. 
For recent reviews see Refs.~\cite{Mosel:2016cwa,Katori:2016yel,Alvarez-Ruso:2017oui}.}
\eqs{
\nu_\ell + n &\to \ell^- + p \, \\
\bar{\nu}_\ell + p &\to \ell^+ + n \,. 
}
Neglecting power corrections to four-fermion theory of order $Q^2/M_W^2$ 
($M_W$ is the $W^\pm$ boson mass), 
the cross section in the laboratory frame is%
\footnote{Our sign convention assumes negative axial charge $F_A(0) \equiv g_A < 0$, hence the negative sign before $B(Q^2)$. For antineutrino-proton scattering, this sign is positive. Our expression corresponds to Ref.~\cite{LlewellynSmith:1971uhs} and differs from Ref.~\cite{Kuzmin:2007kr} in the axial form factor contribution to the function $A$.}
\begin{align}\label{eq:ABC}
\frac{d\sigma}{dQ^2} (Q^2, E_\nu) = \frac{G_F^2 |V_{ud}|^2}{8\pi} \frac{M^2}{E_\nu^2}
\left[ A(Q^2) \frac{m_\ell^2 + Q^2}{M^2} - B(Q^2) \frac{s-u}{M^2} + C(Q^2) \left( \frac{s-u}{M^2} \right)^2 \right] \,,
\end{align}
where $G_F$ is the Fermi constant, $V_{ud}$ is a Cabibbo--Kobayashi--Maskawa (CKM) matrix element, $M = (M_p + M_n)/2$ is the average nucleon mass, $m_\ell$ is the final-state lepton mass, $E_\nu$ is the incoming neutrino energy, and the difference in Mandelstam variables can be written as $s-u = 4E_\nu M - Q^2 - m_\ell^2$.
The three structure-dependent functions $A,~B$, and $C$ are given by
\eqsl{ABC}{
A &= 2\tau_N (F_1^V + F_2^V)^2 - (1+\tau_N) \Big[ (F_1^V)^2 + \tau_N (F_2^V)^2 - (F_A)^2 \Big] 
\\ &
\quad - r_\ell^2 \left[ (F_1^V + F_2^V)^2 
+ \left( F_A  + 2 F_P \right)^2 
- 4(1+\tau_N) F_P^2\right] \,, \\
B &= 4\tau_N F_A (F_1^V + F_2^V) \,, \\
C &= \frac14 \left[ (F_1^V)^2 + \tau_N (F_2^V)^2 + (F_A)^2 \right] \,,
}
where $\tau_N = Q^2/(4M^2), r_\ell = m_\ell/(2M)$, and the four form factors $F_1^V, F_2^V, F_A, F_P$ are defined by
\begin{equation}\label{eq:JW}
\langle p(p^\prime) | \bar{u} \gamma^\mu P_L d | n(p) \rangle = \frac12 \bar{u}^{(p)}(p^\prime)\! \left[ 
\gamma^\mu F_1^V(Q^2) + \frac{i\sigma^{\mu\nu} q_\nu}{2M} F_2^V(Q^2) 
+ \gamma^\mu \gamma_5 F_A(Q^2) + \frac{q^\mu}{M} \gamma_5 F_P(Q^2) \right]\! u^{(n)}(p) \,,
\end{equation}
with $P_L = (1-\gamma_5)/2$.
Equation~(\ref{eq:ABC}) represents the ``Born" cross section for the quasielastic process, 
analogous to Eq.~(\ref{eq:Born}) for the case of $ep$ scattering.  
Soft radiation effects and two-boson exchange contributions have been subtracted
and are to be treated separately.  
It is important to include such radiative corrections and to account for collinear and hard-photon emission in a practical experiment~\cite{DeRujula:1979grv,Day:2012gb,Graczyk:2013fha,Hill:2016gdf}; 
however, our focus here is to determine the Born cross section for the quasielastic process. 

The axial form factor $F_A$ is taken from Ref.~\cite{Meyer:2016oeg}.
In order to illustrate the utility of the new vector form factors, we will use the standard partially-conserved axial current (PCAC) ansatz for the pseudoscalar form factor $F_P$ (whose effects are suppressed by powers of the lepton mass), 
\begin{align}
F_P(Q^2) &= \frac{2M^2}{m_\pi^2 + Q^2} F_A(Q^2) \,.
\end{align}
The isovector vector form factors $F_1^V$ and $F_2^V$ are determined either by 
taking the difference of proton and neutron form factors, or by directly implementing 
the isospin-decomposed fit.  
We have ignored second-class form factors in Eq.~(\ref{eq:JW}), 
and isospin-violating corrections to the relation of $F_i^{p,n}$ to 
$F_i^{V}$.  These effects are suppressed by the fine structure constant $\alpha$ or by $(m_d-m_u)/\Lambda_{\rm QCD}$, and are expected to be small compared to other uncertainties. 

\begin{figure}[tb!]
\centering
\includegraphics[scale=0.65]{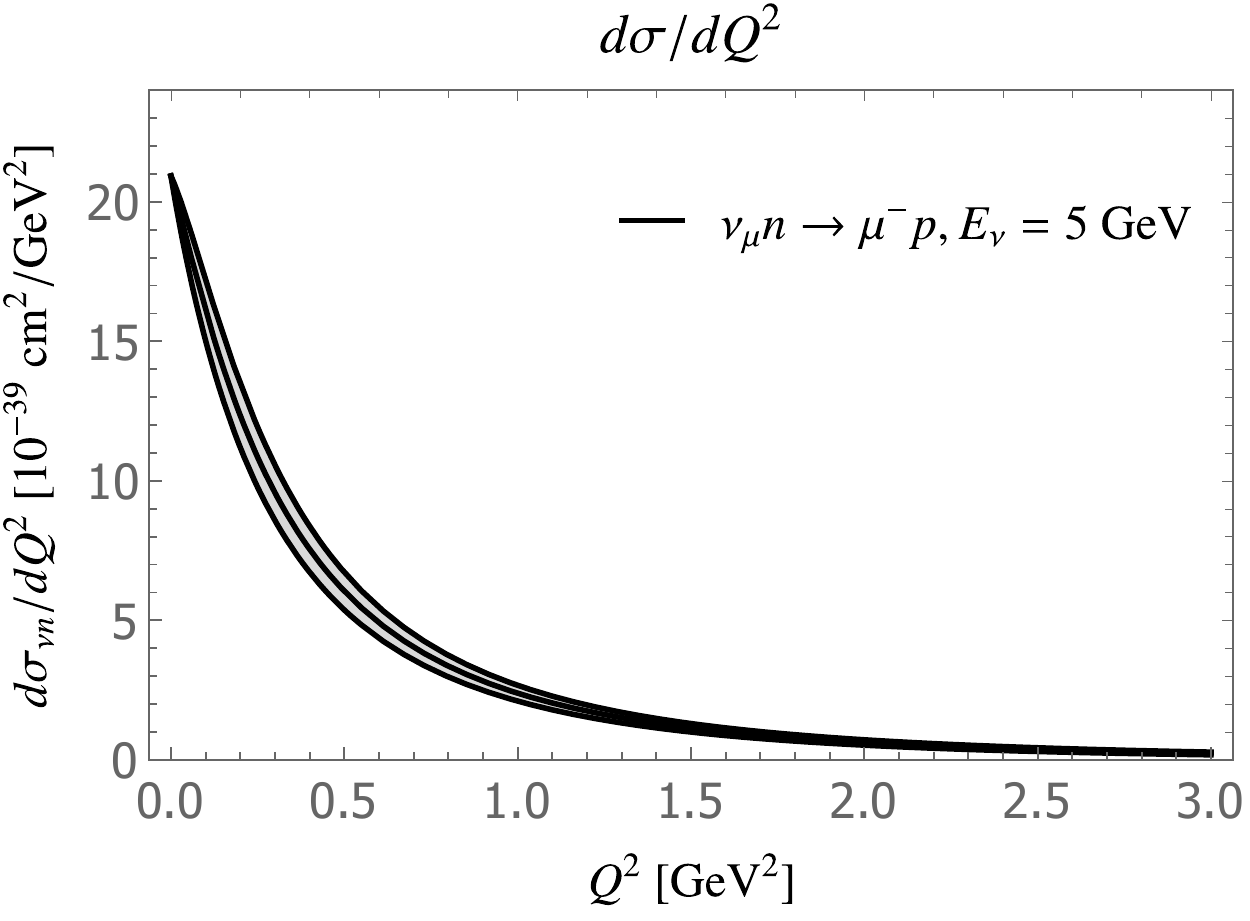}
\includegraphics[scale=0.65]{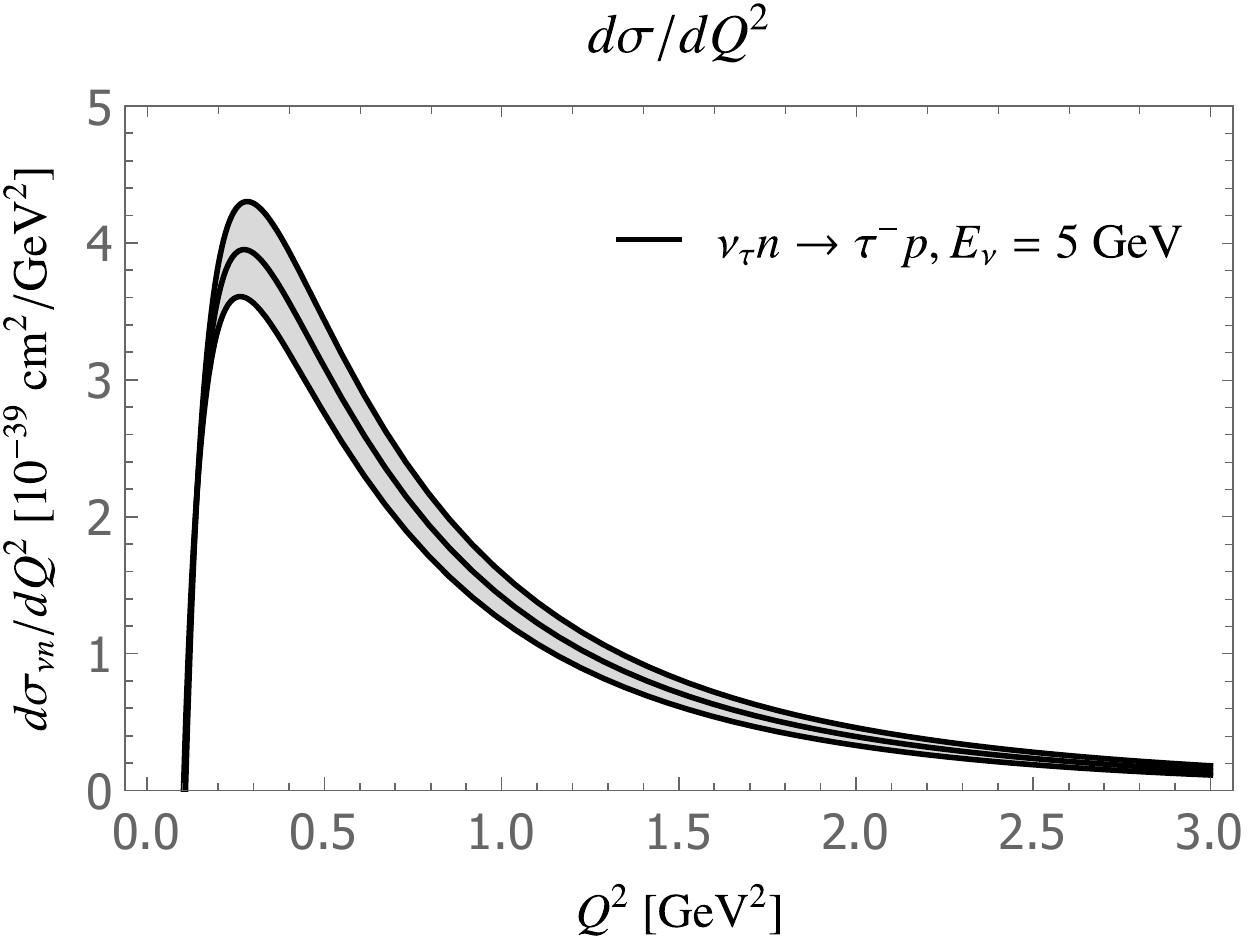}
\caption{Neutrino--neutron quasielastic scattering cross section versus $Q^2$ for muon [left] and tau [right] flavors, using the iso (1~GeV$^2$) fit.}
\label{fig:dsigmadQ2}
\end{figure}

\begin{figure}[htb!]
\centering
\includegraphics[scale=0.65]{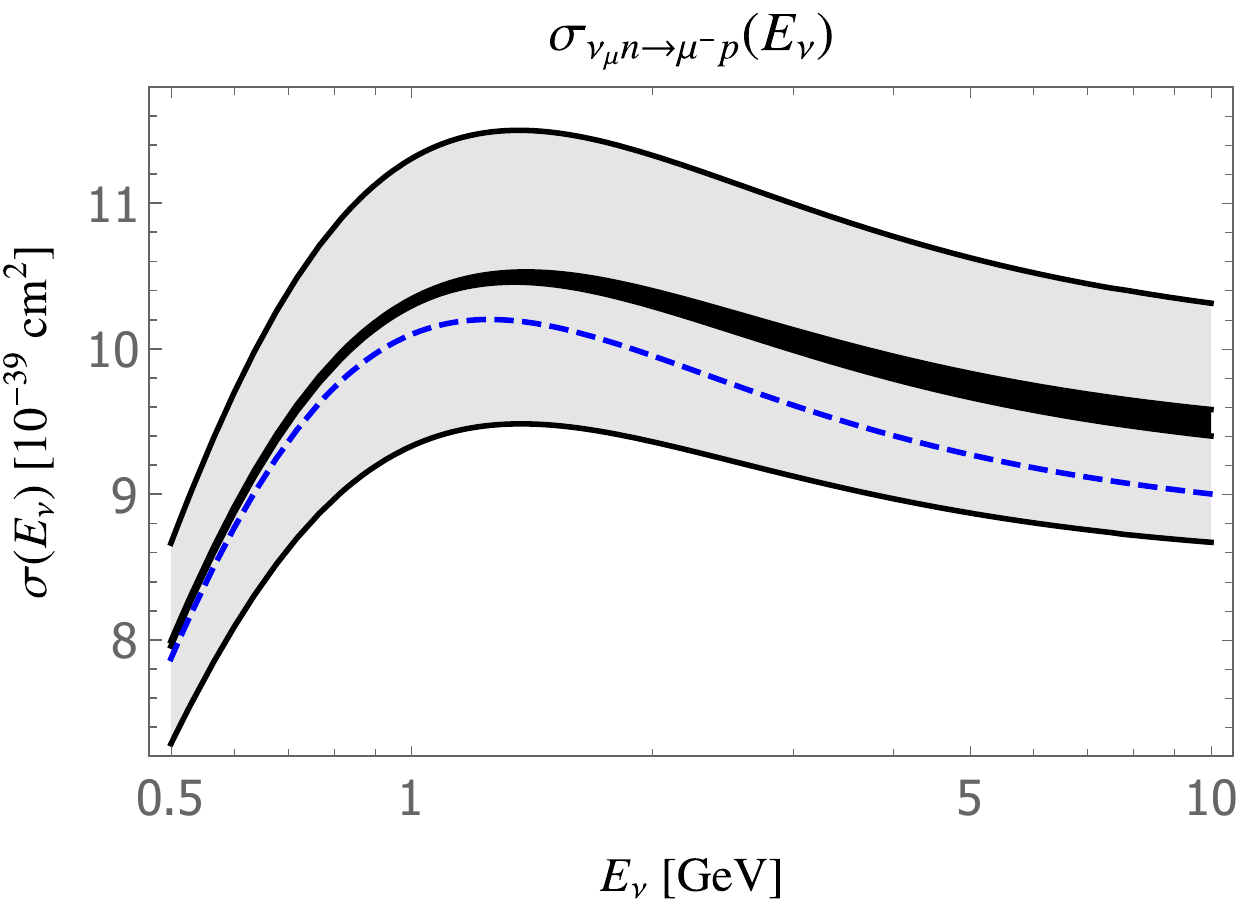}
\includegraphics[scale=0.65]{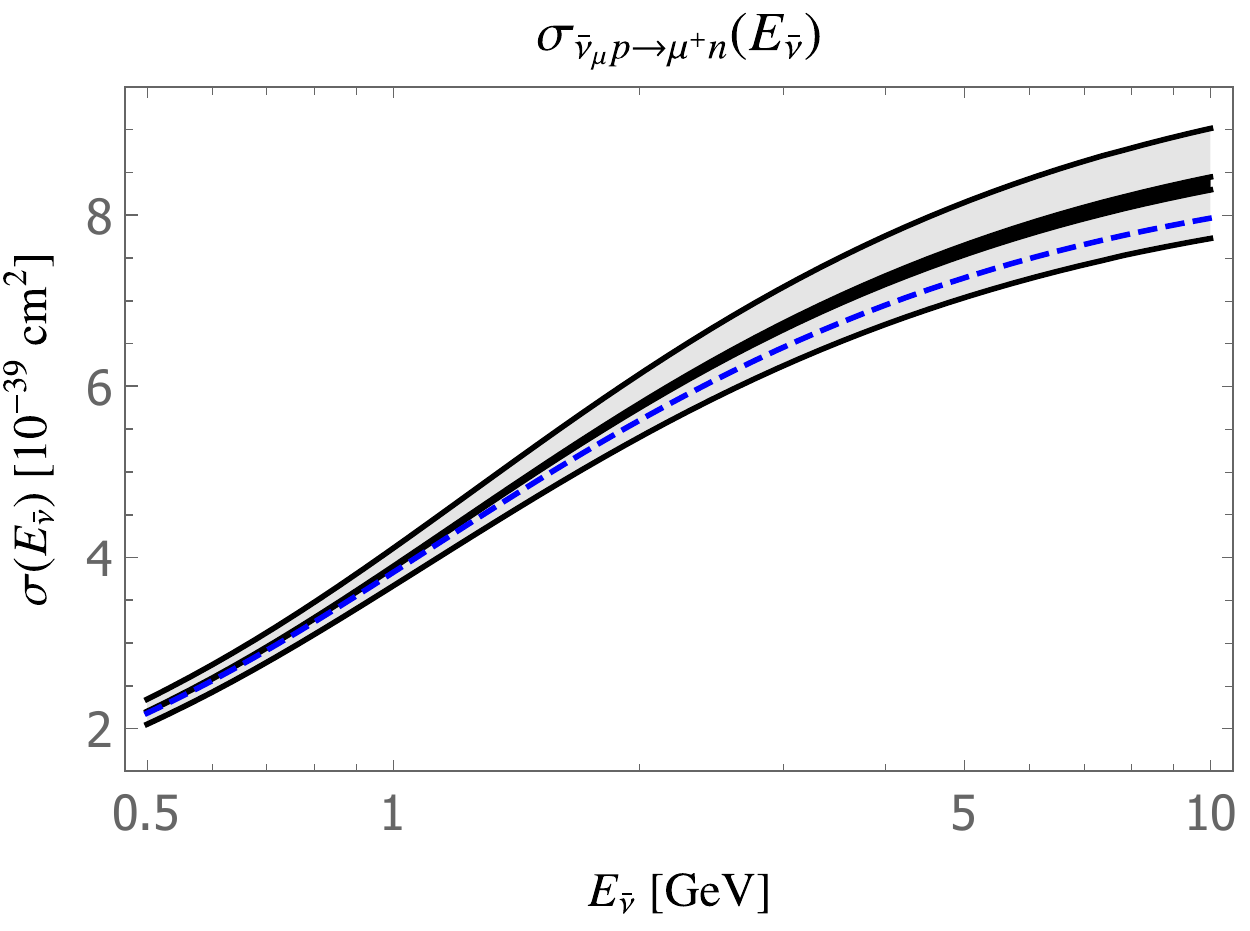}
\caption{Muon neutrino--neutron [left] and antineutrino--proton [right] quasielastic cross section.
Our result is given by the narrow dark band representing 
the iso (1~GeV$^2$) fit with vector form factor 
uncertainty.  Axial form factor uncertainty is represented by the 
wide light band and is to be added in quadrature.  The blue dashed 
line represents the central value using the same axial form factor
as the central curve, but BBBA2005 vector form factors.}
\label{fig:nuNxs1}
\end{figure}

\begin{figure}[htb!]
\centering
\includegraphics[scale=0.65]{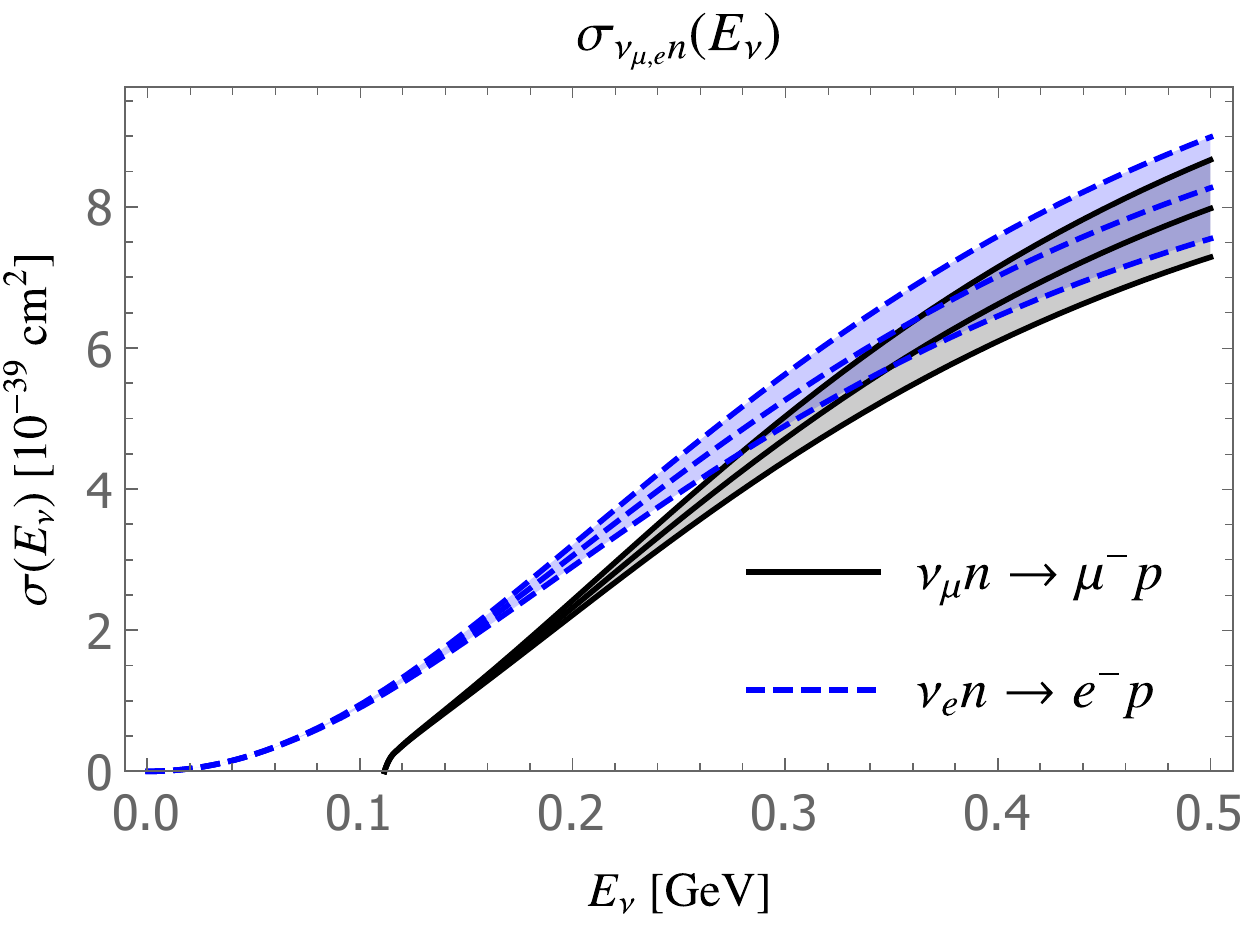}
\includegraphics[scale=0.67]{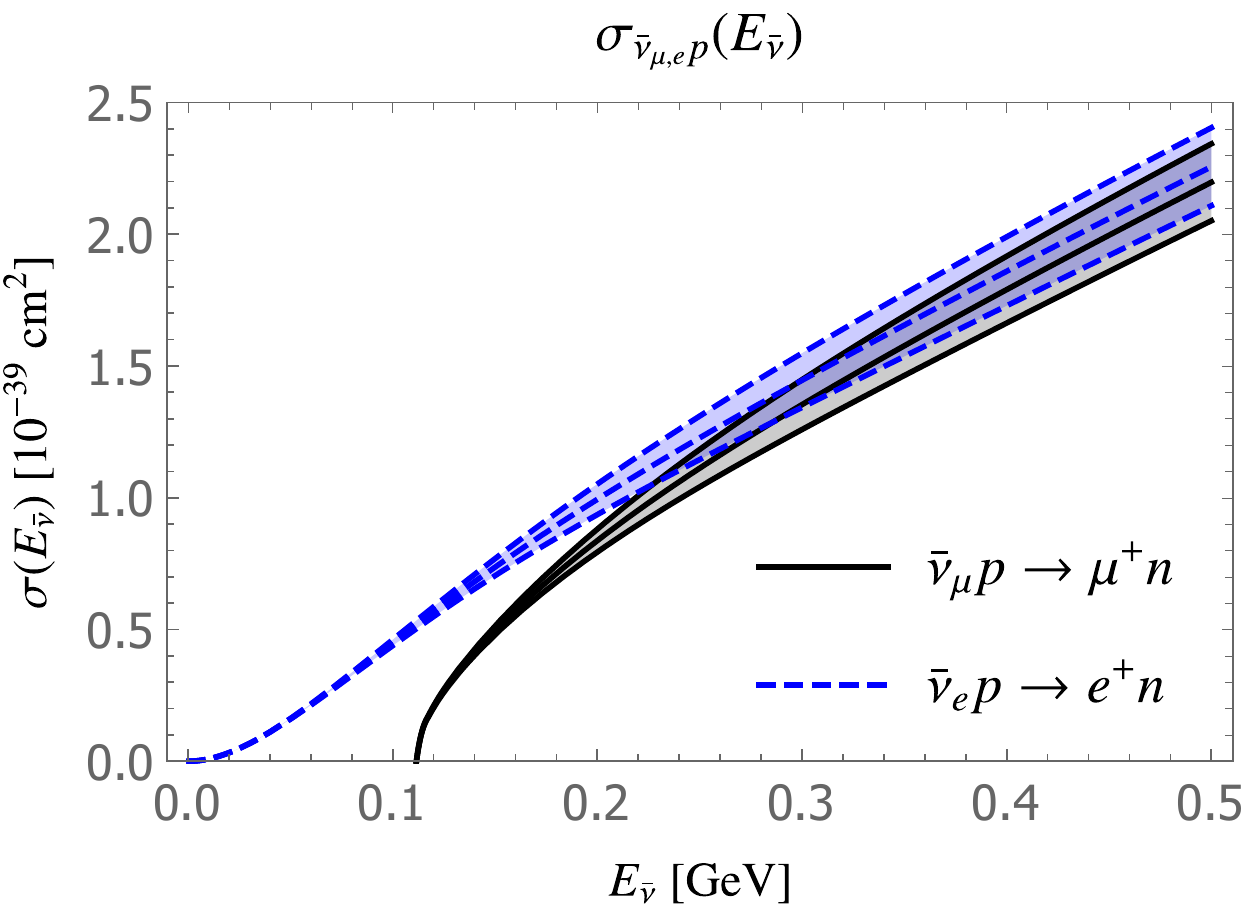}
\caption{Electron and muon neutrino--neutron [left] and antineutrino--proton [right] quasielastic cross section at low energies.
The shaded regions correspond to the light bands in Fig.~\ref{fig:nuNxs1}.
The region with solid boundary line represents the $\nu_\mu$ case and the region with dotted boundary line represents
the $\nu_e$ case.
The vector form factor uncertainty from our fit is not resolved in the plots. 
}
\label{fig:nuNxslowE}
\end{figure}

To illustrate the relevant range of $Q^2$ for neutrino beams 
in the GeV energy regime, we display the $\nu_\mu n$ charged-current quasielastic (CCQE) cross section 
as a function of $Q^2$, fixing $E_\nu = 5~\mathrm{GeV}$, in the left-hand side of Fig.~\ref{fig:dsigmadQ2}. 
The cross section is dominated by $Q^2\lesssim 1$~GeV${}^2$, 
and is relatively insensitive to the detailed 
form factor behavior at larger momentum transfers. 
For comparison, the right-hand side of Fig.~\ref{fig:dsigmadQ2}
shows the $\nu_\tau n$ CCQE cross section; 
this rare process accesses somewhat larger $Q^2$.  
In both cases, there can be residual sensitivity to higher-order coefficients in the $z$ expansion that are poorly constrained by the chosen electron scattering dataset. 
This sensitivity can be determined in practice 
by recomputing observables using different values of $k_{\rm max}$.  

Our CCQE cross sections for 
muon (anti)neutrino are displayed as a function of 
neutrino energy in \figref{nuNxs1}, 
using our default isospin-decomposed (iso~1~GeV$^{2}$) fit. 
The current large uncertainty of the axial form factor
dominates the error budget.  
We remark that the deviation
of central values between our fit and the commonly used BBBA2005 model~\cite{bradford06} 
is sizable compared to the axial form factor uncertainty 
at $E_\nu \gtrsim 1\,{\rm GeV}$. 
The cross section depends strongly on lepton flavor
at energies near or below the muon-production threshold, 
as shown in \figref{nuNxslowE}.  

\begin{table}[htb!]
\caption{CCQE cross section at $E_\nu = 0.5$~GeV.
Errors are from axial form factor $(A)$, vector form factors [$(V)$ for isospin-decomposed fits or $(p)$ and $(n)$ for separate proton and neutron fits], and $z$-expansion truncation ($t$).}
\vspace{3mm}
\centering
\label{Cross_section_0-5GeV}
\begin{minipage}{\linewidth}  
\centering
\begin{tabular}{|l|l|l|}
  \hline          
  Fit choice & $\sigma_{\nu_\mu n \xrightarrow{} \mu^- p}$ $[10^{-39} \, \mathrm{cm^2}]$ & $\sigma_{\Bar{\nu}_\mu p \xrightarrow{} \mu^+ n}$ $[10^{-39} \, \mathrm{cm^2}]$  
\\
\hline
$p, n$ & $7.971(689)_A(7)_p(16)_n (0.2)_t$
& $2.196(146)_A(2)_p(5)_n(0.3)_t$
\\
iso~(1 GeV${}^2$) & $7.975(689)_A(17)_V (1)_t$
& $2.197(146)_A(4)_V(0.5)_t$
\\
iso~(3 GeV${}^2$) & $7.958(689)_A(15)_V (7)_t$
& $2.186(146)_A(4)_V(0.2)_t$
\\
BBBA & $7.87(69)_A(8)_V$ &  $2.18(15)_A(8)_V$ \\
\hline 
\end{tabular}
\end{minipage}
\caption{Same as for Table~\ref{Cross_section_0-5GeV}, but for $E_\nu = 1$~GeV.}
\vspace{3mm}
\centering
\label{Cross_section_1GeV}
\begin{minipage}{\linewidth}  
\centering
\begin{tabular}{|l|l|l|}
  \hline          
  Fit choice & $\sigma_{\nu_\mu n \xrightarrow{} \mu^- p}$ $[10^{-39} \, \mathrm{cm^2}]$ & $\sigma_{\Bar{\nu}_\mu p \xrightarrow{} \mu^+ n}$ $[10^{-39} \, \mathrm{cm^2}]$  
\\
\hline
$p, n$ & $10.312(987)_A(11)_p(22)_n (5)_t$
& $3.886(220)_A(5)_p(8)_n(3)_t$
\\
iso~(1 GeV${}^2$) & $10.319(988)_A(24)_V (6)_t$
& $3.887(220)_A(9)_V(3)_t$
\\
iso~(3 GeV${}^2$) & $10.200(981)_A(20)_V (3)_t$
& $3.851(225)_A(7)_V(1)_t$
\\
BBBA & $10.10(98)_A(18)_V$ &  $3.82(23)_A(8)_V$ \\
\hline
Ref.~\cite{Meyer:2016oeg} & $10.1(9)$ &  $3.83(23)$ \\
\hline 
\end{tabular}
\end{minipage}
\caption{Same as for Table~\ref{Cross_section_0-5GeV}, but for $E_\nu = 3$~GeV.}
\vspace{3mm}
\centering
\label{Cross_section_3GeV}
\begin{minipage}{\linewidth}  
\centering
\begin{tabular}{|l|l|l|}
  \hline          
  Fit choice &$\sigma_{\nu_\mu n \xrightarrow{} \mu^- p}$ $[10^{-39} \, \mathrm{cm^2}]$& $\sigma_{\Bar{\nu}_\mu p \xrightarrow{} \mu^+ n}$ $[10^{-39} \, \mathrm{cm^2}]$  
\\
\hline
$p, n$ & 
$10.035(935)_A(31)_p(66)_n(69)_t$
& 
$6.686(461)_A(19)_p(34)_n(38)_t$
\\
iso~(1 GeV${}^2$) & 
$10.061(936)_A(71)_V(77)_t$
& 
$6.699(460)_A(39)_V(43)_t$
\\
iso~(3 GeV${}^2$) & 
\,\,\,$9.710(918)_A (19)_V (3)_t$
& $6.515(471)_A(13)_V(2)_t$
\\
BBBA & \,\,\,$9.61(91)_A(24)_V$ &  $6.45(47)_A(15)_V$ \\
\hline
Ref.~\cite{Meyer:2016oeg} & \,\,\,$9.6(9)$ &  $6.47(47)$ \\
\hline 
\end{tabular}
\end{minipage}
\end{table}

Tables~\ref{Cross_section_0-5GeV}, \ref{Cross_section_1GeV} and \ref{Cross_section_3GeV} show the CCQE total cross sections 
for three benchmark points with $E_\nu = 0.5, 1$, and $3$~GeV.
In addition to axial and vector form factor uncertainties, we include a $z$-expansion truncation uncertainty estimated from the shift in central value 
when the default fit with $k_{\rm max}=8$ is replaced by the fit with $k_{\rm max}=9$. 
We also compare our evaluation with Ref.~\cite{Meyer:2016oeg}, 
where the BBBA2005 parameterization was used for vector form factors. 

\subsection{Spectroscopy of electronic and muonic atoms \label{sec:atom}}

Modern spectroscopy experiments with ordinary and muonic hydrogen~\cite{Beyer:2017gug,Fleurbaey:2018fih,Bezginov:2019mdi,pohl10,antognini13,Pohl:2016tqq,Adamczak:2016pdb,Ma:2016etb} 
are sensitive to the internal structure of the proton.
In particular, the small size of muonic atoms enhances sensitivity to structure-dependent effects and 
makes measurements with muons attractive in searches for new physics and precise studies of proton and nuclear dynamics.
The leading structure-dependent effect, which is proportional to $\langle r^2 \rangle_E$, 
shifts energy levels at order $m_\ell \alpha^4$ and enters via the exchange of one virtual photon between the lepton ($\ell=e$ or $\ell=\mu$) and the proton.
This effect does not depend on the spin state of the energy level. 
The leading spin-dependent contribution of order $m_\ell \alpha^5$ arises 
from the two-photon exchange. 
It contributes to the hyperfine splitting of energy levels~\cite{antognini13}.
Modern measurements of the Lamb shift in muonic hydrogen~\cite{pohl10,antognini13}, of the hydrogen-deuterium isotope shift~\cite{Parthey:2010aya} and of the $1S$--$2S$ transition in hydrogen~\cite{Parthey:2011lfa} 
are sensitive to spin-independent two-photon exchange contributions as well. 
For both ordinary and muonic hydrogen, the bulk of the two-photon exchange contributions is determined by certain structure parameters, ``moments", 
expressed as $Q^2$ integrals over products of elastic form factors.   
In this section, we compute the Friar and Zemach radii governing spin-independent and spin-dependent 
two-photon exchange, respectively.
Some previous results are compiled in Appendix~\ref{app:complit}.

\subsubsection{Lamb shift}

The leading structure-dependent contribution to the Lamb shift in hydrogen is proportional to the 
(cube of the)
Friar radius $r_F$:
\begin{align}
	r_F^3 = 
	\frac{24}{\pi} \int_0^\infty \frac{dQ^2 }{ Q^5 } \left[ G_E^2(Q^2) - 1 - 2 Q^2 G_E'(0) \right] \,,
\end{align}
where $G_E^\prime(0)=dG_E/dQ^2|_{Q^2=0}$. 
We evaluate $r_F^3$ exploiting the fit of proton and neutron data as well as isospin-decomposed fits and present our results in Table~\ref{Friar_radii0}.
The first error is from the extracted form factor covariance matrix, and the second
is the shift in central value when the default fit with $k_{\rm max}=8$ 
is replaced by the fit with $k_{\rm max}=9$. 
We note that removing the
$\mu$H constraint from our default proton fit shifts 
$(r^p_F)^3 = 2.246(58)\,{\rm fm}^3 \to 2.97(35)\,{\rm fm}^3$. 

\begin{table}[ht]
\centering
\caption{Friar radii of proton and neutron.  The first error is from the extracted form factor covariance matrix and the second error is from $z$-expansion truncation.}
\label{Friar_radii0}
\vspace{3mm}
\begin{minipage}{\linewidth}  
\centering
\begin{tabular}{|l|l|l|c|}
  \hline          
  Fit choice &$\pr{r_F^p}^3~[\mathrm{fm}^3]$&$ \pr{r_F^n}^3~[\mathrm{fm}^3]$ 
\\
\hline
$p, n$ & 
$2.246(58)(2)$
& 
$0.0093(11)(1)$
\\
iso~(1 GeV${}^2$)  &
$2.278(49)(12)$
&
$0.0093(6)(1)$
\\
iso~(3 GeV${}^2$)  &
$2.176(38)(10)$
&
$0.0100(5)(0)$
\\
\hline 
\end{tabular}
\end{minipage}
\end{table}

\subsubsection{Hyperfine splitting}

The first measurements of the $1S$ hyperfine splitting in muonic hydrogen with ppm precision are being planned by the 
CREMA~\cite{Pohl:2016tqq} and FAMU~\cite{Adamczak:2016pdb} Collaborations, and at J-PARC~\cite{Ma:2016etb}.
The leading nucleon-structure contribution to the hyperfine splitting of $S$ energy levels is given by the two-photon exchange diagram. 
The bulk of the correction is proportional to the Zemach radius $r_Z$~\cite{Zemach:1956zz}, 
which can be expressed as a convolution of nucleon electric and magnetic form factors,
\eql{rZdef}{
r^N_Z = -\frac4{\pi} \int_0^\infty \frac{dQ}{Q^2}
\bigg[ \frac{G^N_M(Q^2) G^N_E(Q^2)  - G^N_M(0) G^N_E(0)}{G^N_M(0)} \bigg] \,.
}

\begin{table}[ht]
\centering
\caption{Same as Table~\ref{Friar_radii0} but for Zemach radii.}
\label{Zemach_radii0}
\vspace{3mm}
\begin{minipage}{\linewidth}  
\centering
\begin{tabular}{|l|l|l|c|}
  \hline          
  Fit choice &$ \quad r^p_Z~[\mathrm{fm}]$&$ \quad r^n_Z~[\mathrm{fm}]$ 
\\
\hline
$p, n$ & 
$1.0227(94)(51)$
& 
$-0.0443(26)(1)$
\\
iso~(1 GeV${}^2$)  &
$1.0246(84)(40)$
&
$-0.0445(14)(3)$
\\
iso~(3 GeV${}^2$)  &
$1.0450(58)(45)$
&
$-0.0462(12)(0)$
\\
\hline 
\end{tabular}
\end{minipage}
\end{table}

Similar to the Friar radii, 
we present Zemach radii evaluated using the fits from \secref{fitresults} in Table~\ref{Zemach_radii0}.
These results provide a first rigorous error estimate. 
We note that removing the
$\mu$H constraint from our default proton fit shifts 
$r^p_Z = 1.0227(94)\,{\rm fm} \to 1.0426(132)\,{\rm fm}$. 

\section{Summary} \label{sec:summary}

We have presented a compact representation of the proton and neutron
vector form factors in terms of $z$-expansion coefficients,
including central values, errors and correlations.
The results can be used to evaluate both central values and error bars for many derived quantities that are 
sensitive to GeV and sub-GeV momentum transfers. 

In our default fits we employed the following data: 
(i) the high-statistics Mainz dataset for $ep$ cross sections; 
(ii) $en$ elastic scattering data at momentum transfers $Q^2 \leq 1\,{\rm GeV}^2$; and 
(iii) precise external constraints on the proton and neutron electric charge radii.  
We considered two types of fits to these data.  First, we performed 
separate proton and neutron fits, i.e., proton data fit to proton form factors
and neutron data fit to neutron form factors.  Second, we performed a
fit of both proton and neutron data to isospin-decomposed form factors. 
For proton structure observables, there is only a slight reduction in 
uncertainty when the proton fit is replaced by the isospin-decomposed fit; 
for simplicity we use the proton fit as our final result:
$r_M^p=0.739(41)(23)\,{\rm fm}$; 
$\langle r^4 \rangle^p_E = 1.08(28)(5)\,{\rm fm}^4$;
$\langle r^4 \rangle^p_M = -2.0(1.7)(8)\,{\rm fm}^4$;
$(r_F^p)^3 = 2.246(58)(2)\,{\rm fm}^3$;
$r_Z^p = 1.0227(94)(51)\,{\rm fm}$.
For neutron structure observables, the abundance and precision 
of proton data relative to neutron data lead to a significant 
reduction in uncertainty when using the isospin-decomposed fit;
we thus use the isospin-decomposed fit as our final result: 
$r_M^n = 0.776(53)(28)\,{\rm fm}$;
$\langle r^4 \rangle^n_E = -0.33(24)(3)\,{\rm fm}^4$;
$\langle r^4 \rangle^n_M = -2.3(2.1)(1.1)\,{\rm fm}^4$;
$(r_F^n)^3 = 0.0093(6)(1)\,{\rm fm}^3$;
$r_Z^n = -0.0445(14)(3)\,{\rm fm}$.
For the neutrino CCQE cross sections, only the isovector combination of
vector form factors appears, and we thus use the cross section determined from isospin-decomposed form factors as our default result for the Born cross sections: 
$\sigma_{\nu_\mu n \xrightarrow{} \mu^- p}|_{E_\nu=0.5\,{\rm GeV}} = 7.975(689)(17)(1)$,
$\sigma_{\nu_\mu n \xrightarrow{} \mu^- p}|_{E_\nu=1\,{\rm GeV}} = 10.319(988)_A(24)(6)$ and 
$\sigma_{\nu_\mu n \xrightarrow{} \mu^- p}|_{E_\nu=3\,{\rm GeV}} = 10.061(936)_A(71)(77)$.
We present the uncertainty coming from vector form factors and the truncation uncertainty as the last two errors, respectively, in results of this paragraph.

Significant tensions exist between the default dataset and other $ep$ data. 
In Ref.~\cite{Ye:2017gyb}, we quantified this tension as a 
function of $Q^2$.  Without knowing the source of discrepancy
it remains unclear how to rigorously address this tension in the fit, 
and how to propagate it to derived observables.   
In this paper, we bypass this issue by  focusing on the internally consistent default dataset, but present results
for comparison also from the global fit that includes
World and Pol data as in Table~\ref{tab:datapts}.  
This combined iso~(3~GeV$^{2}$) fit is similar to our global fit from 
Ref.~\cite{Ye:2017gyb}; 
the detailed comparison is discussed in \appref{syst}.
The iso~(3~GeV$^{2}$) fit includes more data than our default fit and it is thus 
not surprising that this fit predicts smaller uncertainties in derived 
observables.  
However, the fit does not address internal dataset tensions and the iso~(3~GeV$^{2}$) uncertainties are likely underestimates. 

A primary goal of this work is to provide a consistent framework for applications such as neutrino event generators to propagate form factor 
constraints and uncertainties into cross section predictions.  
The framework is readily adapted to new data.
Our new precise vector form factors have small uncertainty but deviate significantly from 
commonly used parameterizations; such deviations become sizable compared to the dominant axial form factor uncertainty for larger neutrino energies.  
It is important to address these discrepancies with future experimental and/or lattice QCD data. 
We remark that the axial form factor was extracted under a specific (BBBA2005) assumption for the vector form factors.  This ansatz can be justified given the current large uncertainty of elementary target neutrino data.  However, correlations between vector and axial form factors should be accounted for when future more precise data become available. 

\vskip 0.2in
\noindent
{\bf Acknowledgements}
\vskip 0.1in
\noindent
R.J.H. and G.L. thank J.~Arrington and Z.~Ye for many discussions and collaboration 
on Refs.~\cite{Lee:2015jqa,Ye:2017gyb} which inspired the present work.  Similarly, 
R.J.H. thanks M.~Betancourt, R.~Gran and A.~Meyer for discussion and collaboration on 
Ref.~\cite{Meyer:2016oeg}. 
The research of K.B., R.J.H. and O.T. supported by the U.S. Department of Energy, Office of Science, Office of High Energy Physics, under Award No.~DE-SC0019095. 
Fermilab is operated by Fermi Research Alliance,
LLC under Contract No. DE-AC02-07CH11359 with the U.S. Department of Energy. 
G.L. acknowledges support by the Samsung Science \& Technology Foundation under Project No.~SSTF-BA1601-07, 
a Korea University Grant, and the support of the U.S. National Science Foundation through Grant No.~PHY-1719877.
G.L. is grateful to the Technion--Israel Institute of Technology, the Fermilab theory group, and the Mainz Institute for Theoretical Physics (MITP) for hospitality and partial support during completion of this work.
O.T. acknowledges the Fermilab theory group and the theory group of Institute for Nuclear Physics at
Johannes Gutenberg-Universit\"at Mainz for warm hospitality and support. 
The work of O.T. is supported by the Visiting Scholars Award Program of the Universities Research Association.

\appendix

\section{Consistency between datasets} \label{app:syst}

\begin{figure}[htb!]
\centering
\includegraphics[width=0.48\textwidth]{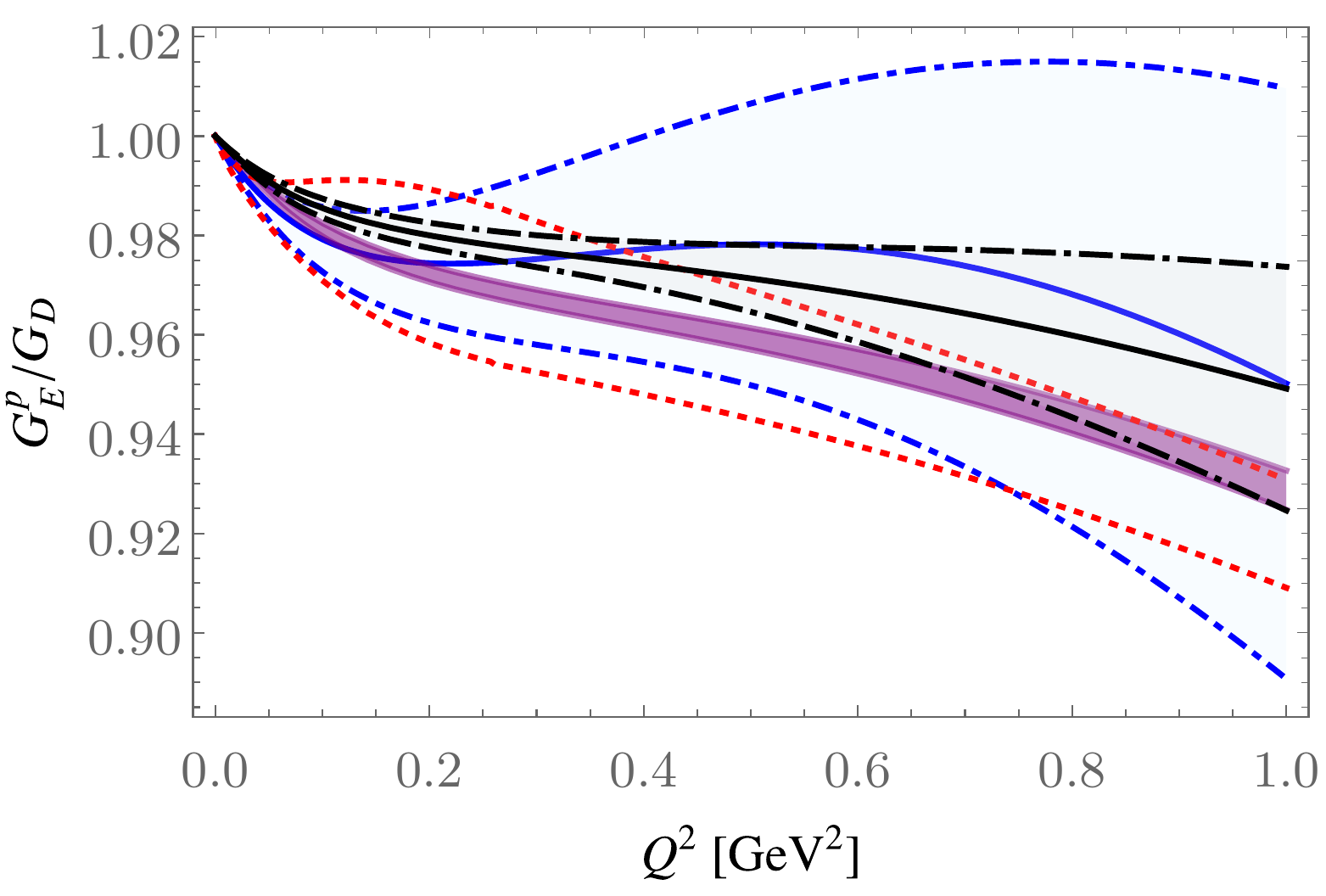}
\includegraphics[width=0.48\textwidth]{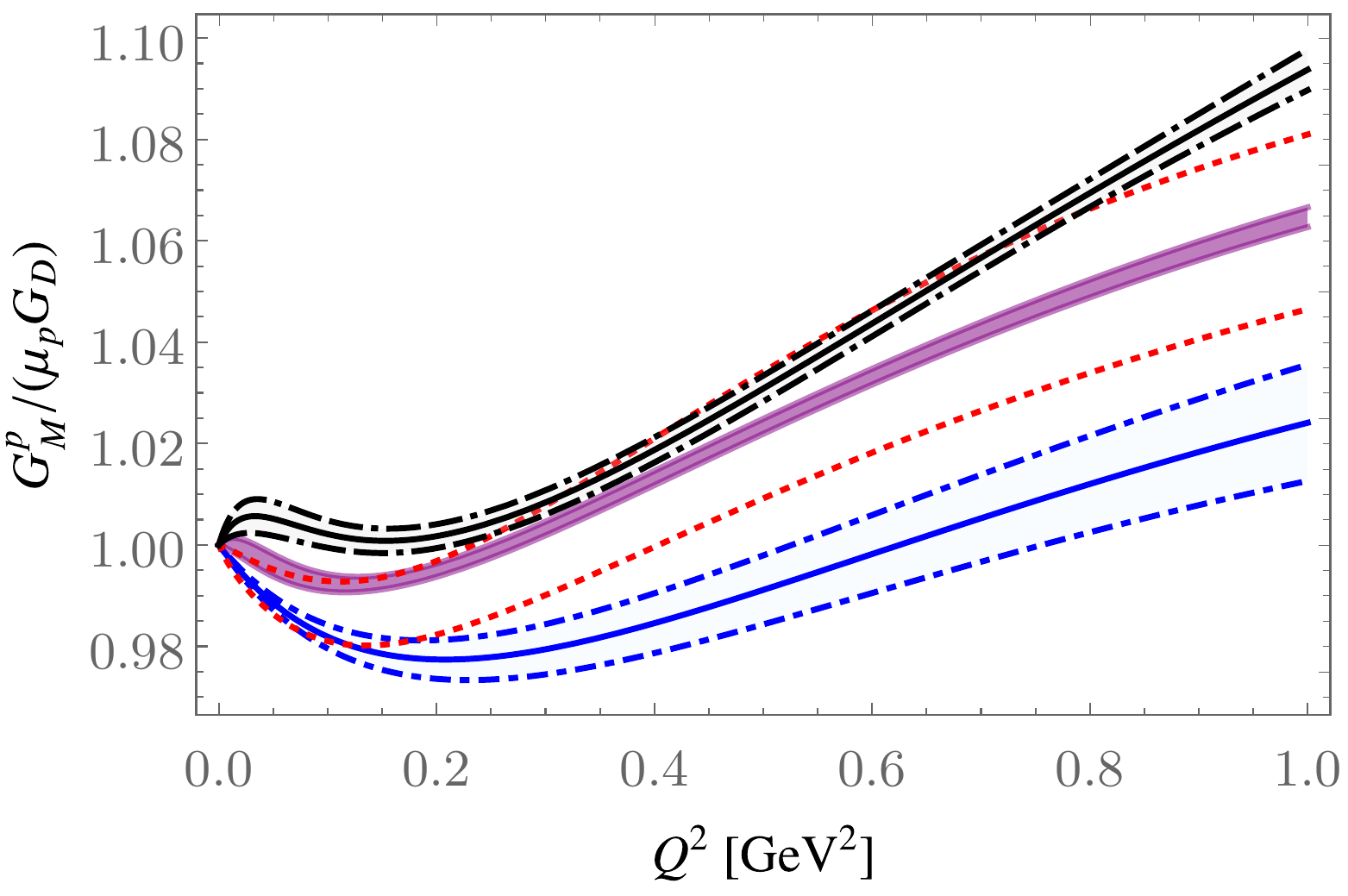}
\includegraphics[width=0.48\textwidth]{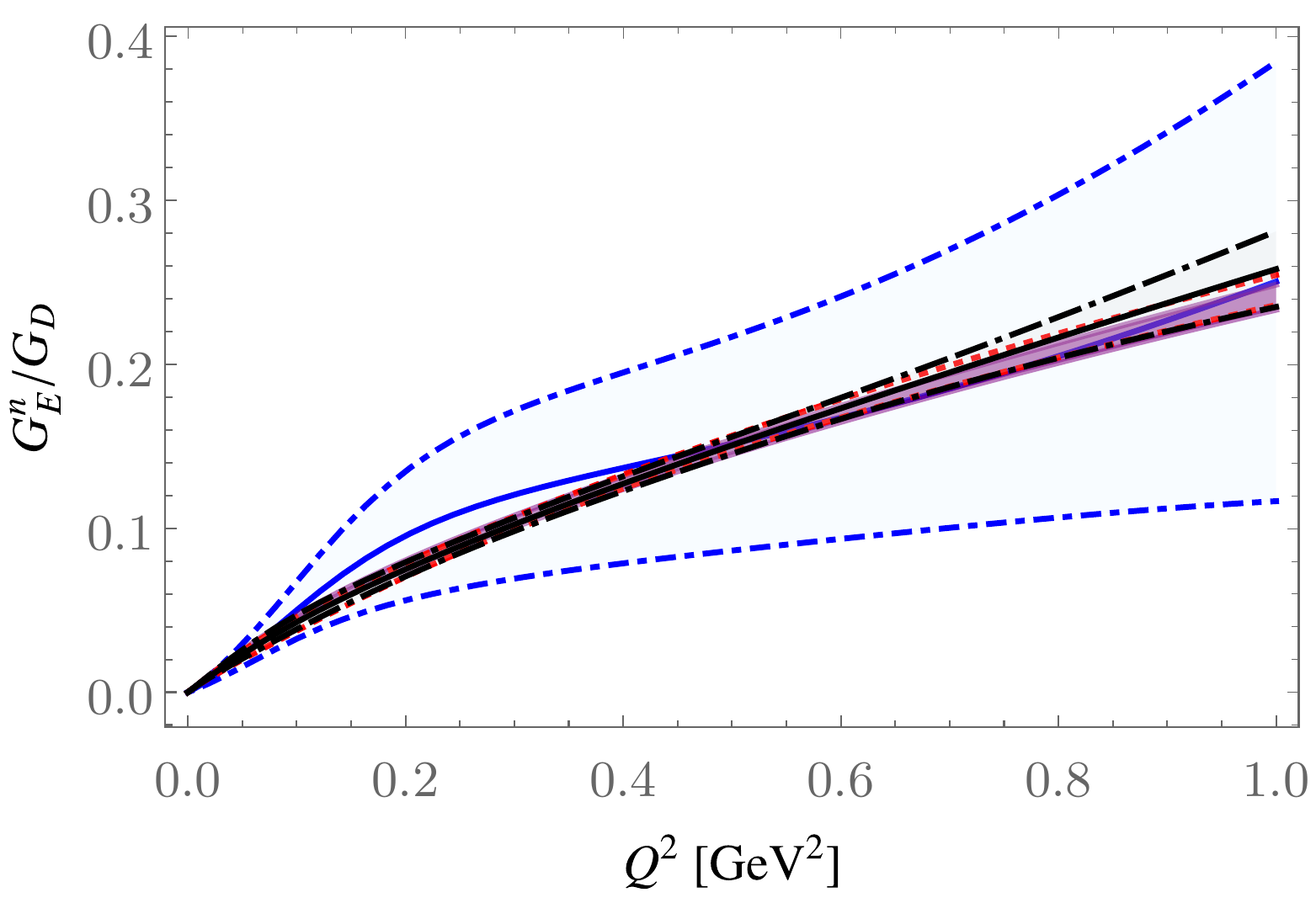}
\includegraphics[width=0.48\textwidth]{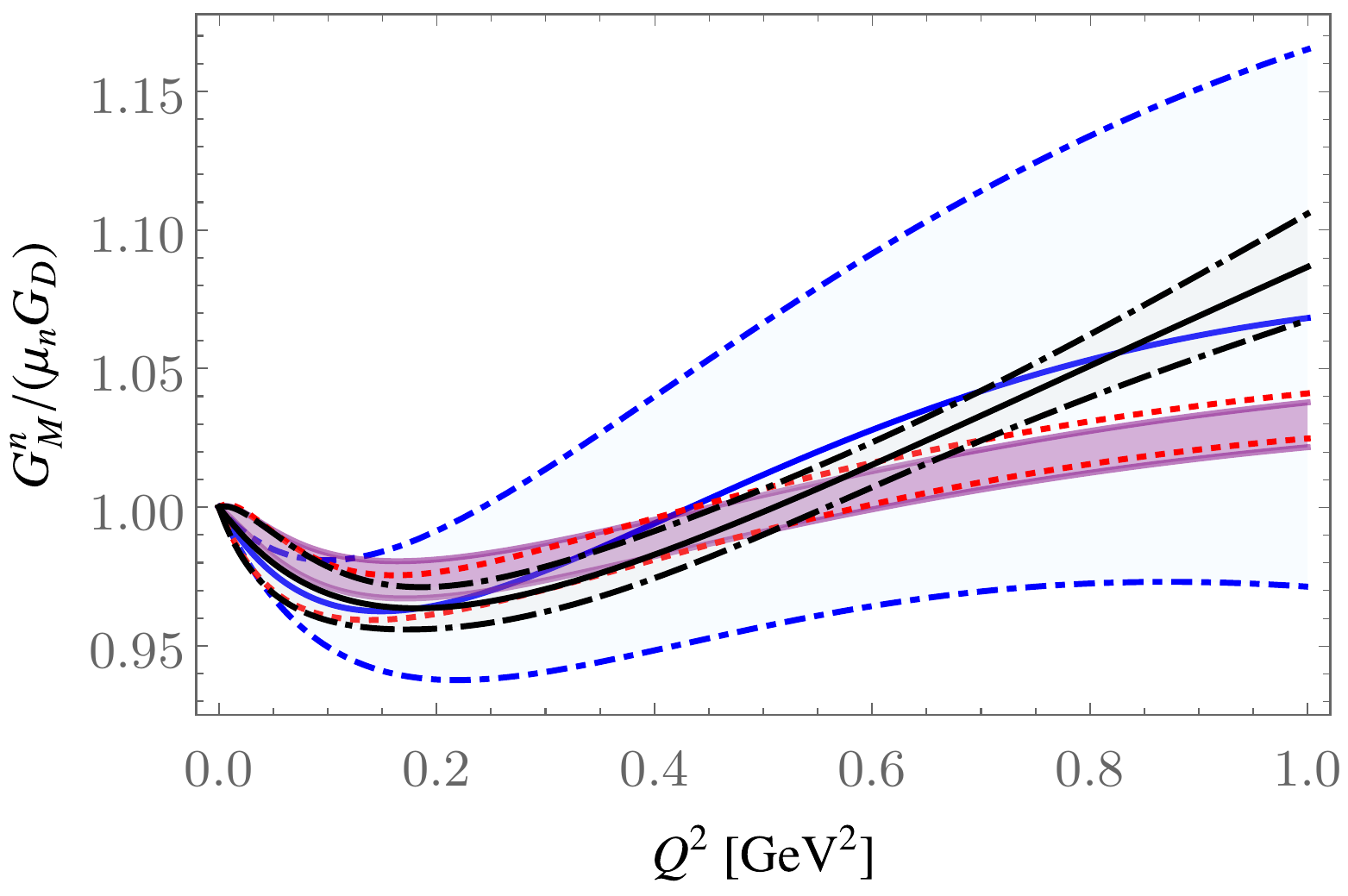}
\includegraphics[width=0.48\textwidth]{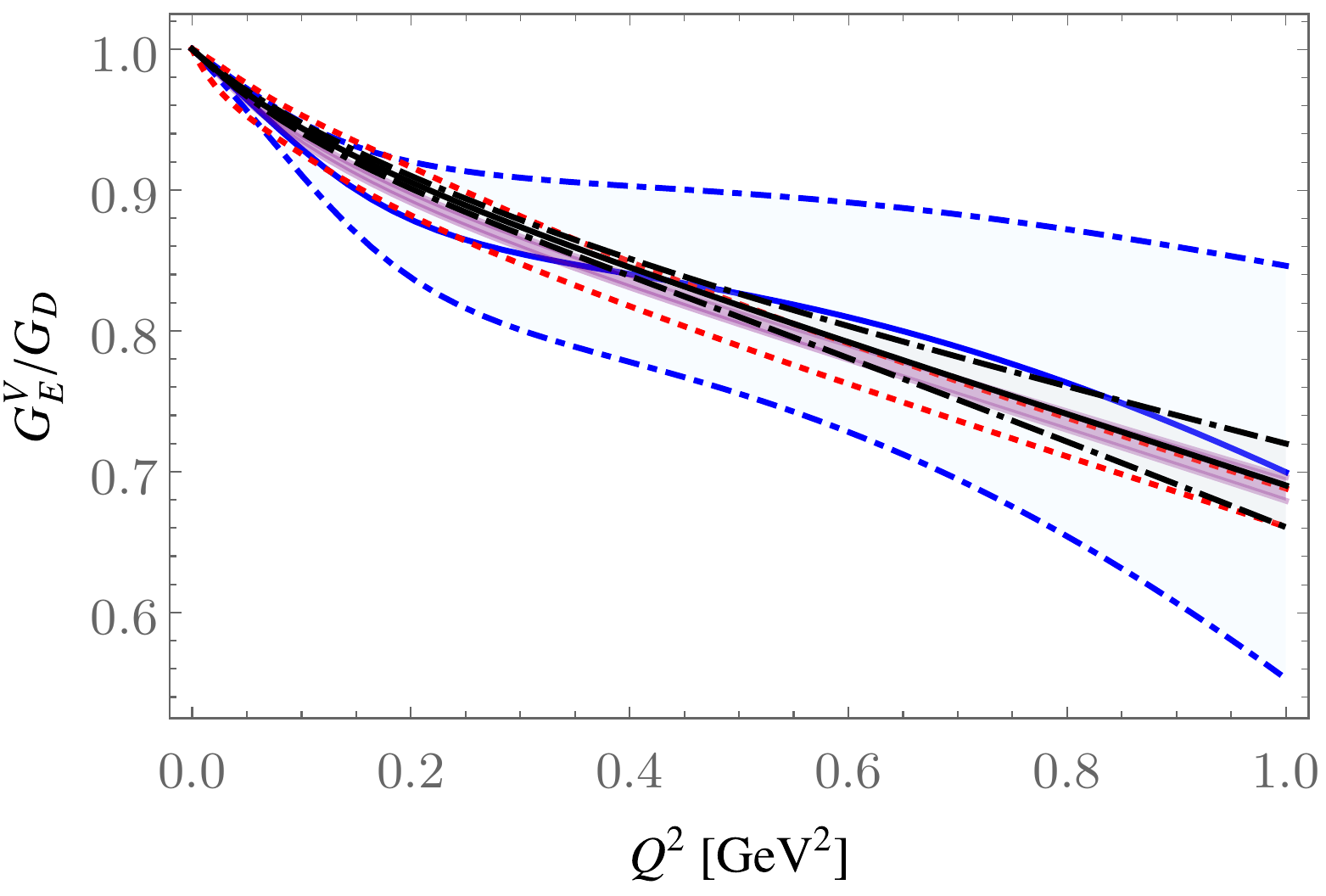}
\includegraphics[width=0.48\textwidth]{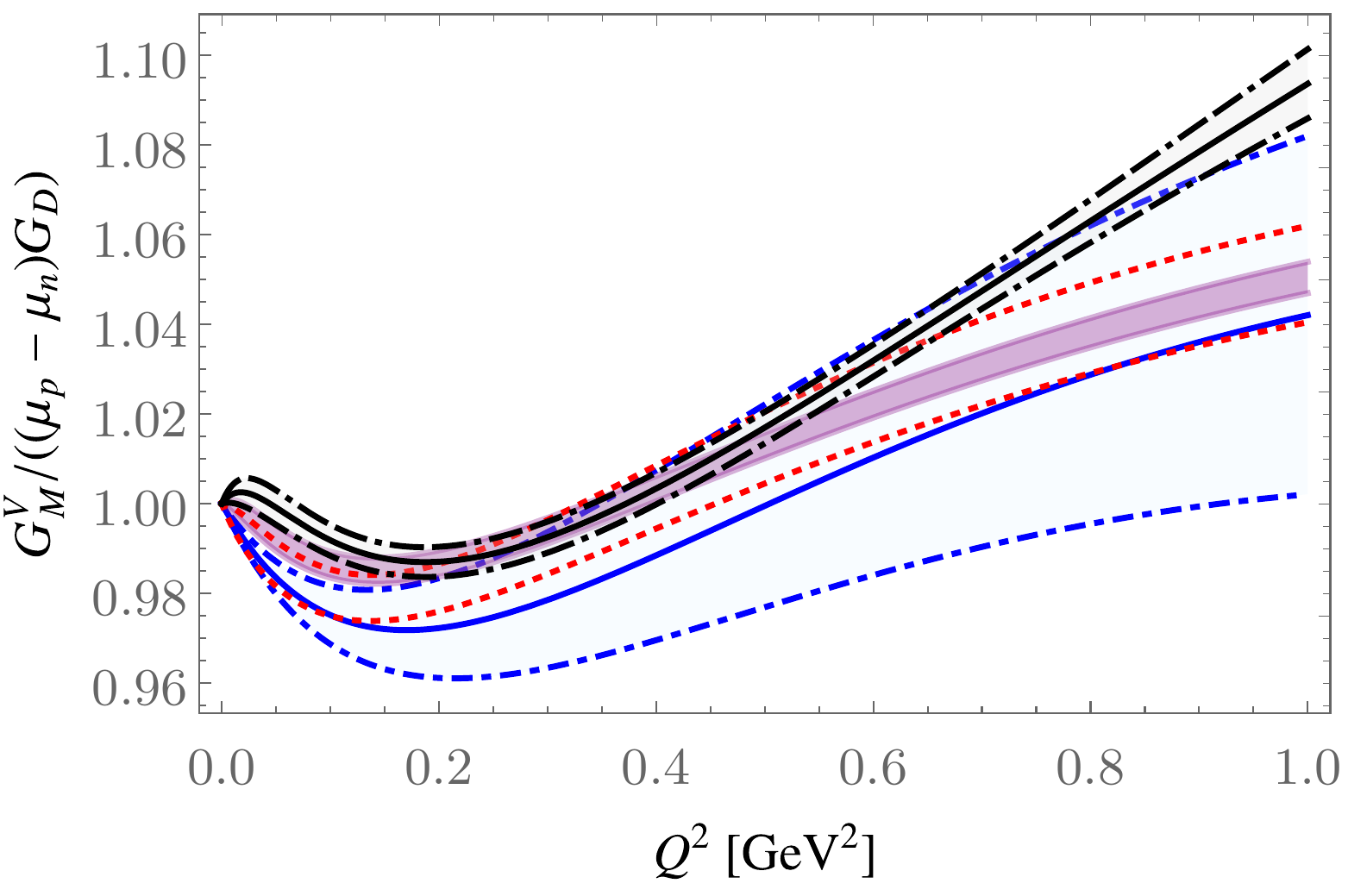}
\caption{Plots of $1\sigma$ bands of 
$G_E^p$ and $G_M^p$ [top], 
$G_E^n$ and $G_M^n$ [middle],
and $G_E^V$ and $G_M^V$ [bottom]
from different fits.
The black long dash-dotted curves are the results of the following: 
the $p$ fit of line 1 in Table~\ref{tab:datapts} [top]; the $n$ fits of lines 2 and 3 in Table~\ref{tab:datapts} [middle]; and 
the iso ($1 \GeV^2$) fit of line 4 in Table~\ref{tab:datapts} [bottom].
The purple bands are the results of the iso ($3 \GeV^2$) fit of line 5 in Table~\ref{tab:datapts}.
The red dotted curves correspond to the global fit of Ref.~\cite{Ye:2017gyb},
and the blue dash-dotted curves are the BBBA2005 result of Ref.~\cite{bradford06}.}
\label{fig:FFcomp}
\end{figure}

In this appendix, we discuss the tension between iso~(1$~\mathrm{GeV^2}$) and iso~(3$~\mathrm{GeV^2}$) fits 
and compare to our previous global fit~\cite{Ye:2017gyb} and 
to the BBBA2005 parameterization~\cite{bradford06}.   
In particular, we illustrate the tension between extractions of 
$G_M^p$ at $Q^2 \sim 1\,{\rm GeV}^2$ from the Mainz and other World cross-section data. 
This tension manifests in observables sensitive to moderate $Q^2 \gtrsim {\rm few}\times 0.1\,{\rm GeV}^2$, such as CCQE cross sections with few GeV neutrino energies; cf. Fig.~\ref{fig:nuNxs1}.
We also show that including the PRad data does not significantly 
alter the  fits when the $\mu$H constraint is imposed. 

\subsection{Mainz and World+Pol datasets} 

Figure~\ref{fig:FFcomp}
compares the form factors from our default $p$, $n$, and iso~(1~GeV${}^2$) fits 
to our iso~(3~GeV${}^2$) fit.
We also compare to our previous global fit from 
Ref.~\cite{Ye:2017gyb}.  
That global fit corresponds with the iso~(3~GeV${}^2$) fit after 
the following modifications: 
(i) inclusion of the $\mu$H constraint (\ref{eqn:rEpmuH});
(ii) improved treatment of TPE correction (\ref{eqn:TPE}); 
(iii) omission of data above $Q^2=3\,{\rm GeV}^2$;
(iv) choice of form factor expansion parameters $t_0$ and $k_{\rm max}$ optimized for 
$0<Q^2<1\,{\rm GeV}^2$. 
Note that the error band from Ref.~\cite{Ye:2017gyb} includes an \textit{ad hoc}
``data tension" error to account for the tension between Mainz and other World data.
Since we have in mind applications to neutrino 
cross sections, we compare also to the commonly used BBBA2005 parameterization. 
The BBBA2005 parameterization resulted from a fit to data preceding the A1 experiment, 
and is in severe tension with our default fit for $G_M^p$.

\subsection{PRad and Mainz datasets}
\label{app:PRAD}

The PRad Collaboration recently presented new measurements of elastic electron--proton scattering 
at JLAB~\cite{Xiong:2019umf}.
At two beam energies $E = 1.1 \GeV$ and $2.2 \GeV$, 33 and 38 measurements were taken in the range of $Q^2$ up to $0.016 \GeV^2$ and  $0.058 \GeV^2$, respectively.
PRad announced a result 
$r_E^p = 0.831 \pm 0.007_{\mathrm{stat}} \pm 0.012_{\mathrm{syst}}$~fm 
fitting to a rational functional form for $G_E^p$,
\eql{rational11}{
G_E^p (Q^2) = \frac{1 + p_1 Q^2}{1 + p_2 Q^2} \,.
}
Notably, this is within $1\sigma$ of $\pr{r_E^p}_{\muH}$ in \eqnref{rEpmuH} 
from muonic hydrogen spectroscopy.

The PRad Collaboration employed particular assumptions in fitting the cross sections, which are detailed in the supplementary material of Ref.~\cite{Xiong:2019umf}. 
To extract form factors, the measured scattering cross sections were fit to the following reduced cross section
\eql{PRAD_ansatz}{
\sigma^{\mrm{red}}_{\mrm{PRad}} = (n G_E)^2 + \frac{\tau}{\varepsilon} (G_M^K)^2 \,,
}
where $\tau=Q^2/(4M_p^2)$, $\epsilon= [1+2(1+\tau)\tan^2(\theta/2)]^{-1}$,
$n$ is a normalization parameter for a given beam energy, 
and $G_M^K$ is the Kelly parameterization for the proton magnetic form factor~\cite{kelly04}.
The PRad Collaboration showed that the cross sections vary by less than 0.2\% when different models for 
the magnetic form factor are used.%
\footnote{
Note that after factoring out the normalization parameter from the reduced cross section, 
the ansatz in \eqnref{PRAD_ansatz} does not strictly reproduce the correct anomalous magnetic moment.  
Since the parameter $\tau$ is small in the range of $Q^2$ covered by the PRad experiment, the fits are insensitive to the replacement $G_M^K \to G_M^K /n$.}

\begin{figure}[t]
\centering
\includegraphics[width=0.49\textwidth]{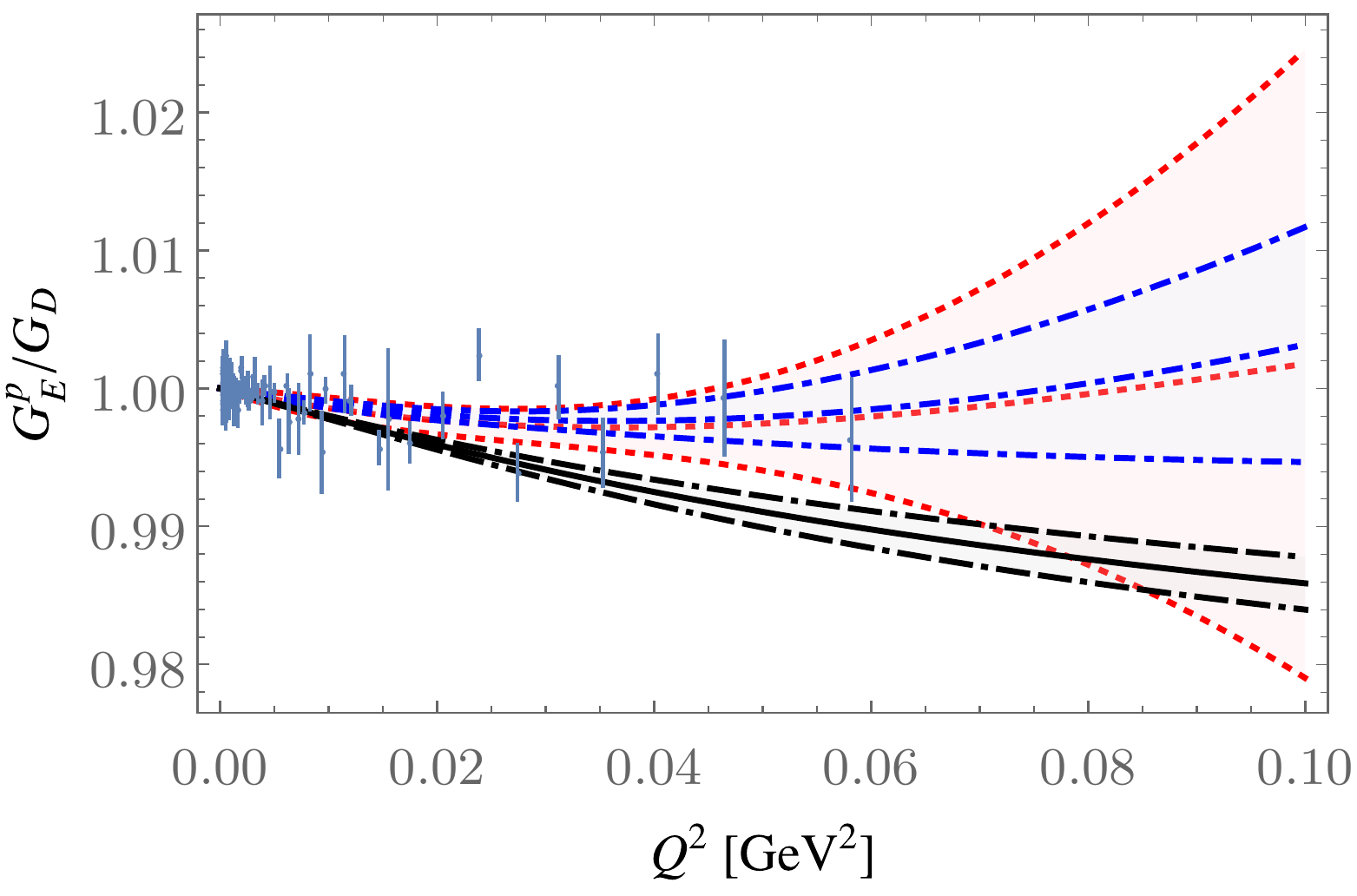}
\includegraphics[width=0.49\textwidth]{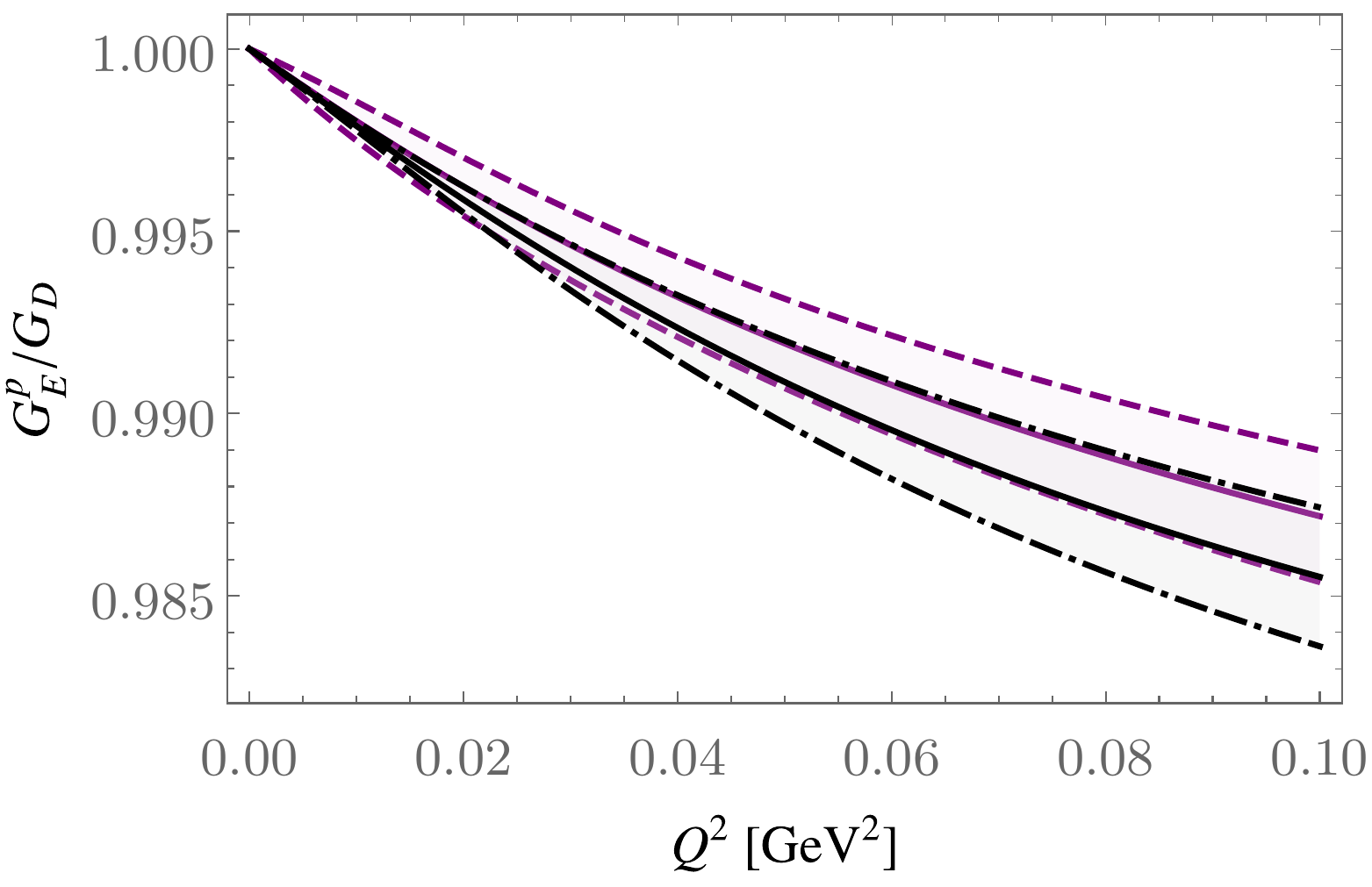}
\caption{Comparison of $G_E^p$ from fits with and without PRad data.
In both plots, the black, long dash-dotted curve is our default (proton) fit.  
On the left-hand side, blue points are the tabulated PRad form factors with statistical errors; the blue, dash-dotted curve is the PRad extraction; and the red, dotted curve is our extraction from PRad data.
On the right-hand side, we compare our default fit to the fit when the $\mu$H constraint is replaced by PRad data (purple, dashed curve).
}
\label{fig:GEpPRad}
\end{figure}

In \figref{GEpPRad}, we compare $G_E^p$ from four fits: 
\begin{enumerate}
    \item (blue, dash-dotted curve in left plot) Fitting with the above rational functional form to the provided form factor tabulations with statistical-only errors, we have reproduced the PRad results for the proton charge radius and reduced $\chi^2$.
    \item (red, dotted curve in left plot) Following a modified version of the PRad procedure, we also fit directly to the tabulated PRad cross sections with statistical and systematic uncertainties added in quadrature.
    Fixing the magnetic form factor to the dipole form $G_M^p(Q^2) = \mu_p G_D(Q^2)$,
    we employed the $z$ expansion (without sum rules) with $k_{\max} = 3$ for $G_E^p$.
    For each beam energy, a separate normalization parameter multiplying the entire reduced cross section is used. We did not apply additional TPE corrections. The extracted radius value is $r_E^p=0.836(19)\,{\rm fm}$, with $\chi^2 = 23.88$ and $n_{\mathrm{dof}} = 68$.  
    \item (black, long dash-dotted curve in both left and right plots) $G_E^p$ obtained from our default proton-only fit to Mainz data with $\pr{r_E^p}_{\muH}$ constraint. 
    \item (purple, solid/dashed curve in right plot) $G_E^p$ obtained from the proton-only fit using the $z$ expansion with $k\ld{max} = 8$ to combined Mainz and PRad data (statistical and systematic uncertainties added in quadrature) without $\pr{r_E^p}_{\muH}$ constraint. 
    TPE corrections are applied to both datasets.
    The extracted radius value for this fit is $r_E^p = 0.843(11)$~fm, with $\chi^2 = 503.24$ and $n_{\mathrm{dof}} = 720$.
\end{enumerate}
In the right plot of Fig.~\ref{fig:GEpPRad}, we show that the combined fit {\it without} 
the $\mu$H constraint and our default fit {\it with} the $\mu$H constraint lie within the $1\sigma$ uncertainty bands of each other for the entire $Q^2$ range of the PRad data.
It would be desirable to include the PRad data directly into our fits, alongside the data 
in Table~\ref{tab:datapts}; 
we refrain from doing so since the PRad uncertainties are
systematics dominated and uncertainty correlations are not yet available~\cite{Xiong:2019umf}.  
We have shown, however, that these data will not significantly alter our fits once the 
precise external $\mu$H constraint is imposed. 
Taking the PRad errors at face value (i.e., neglecting correlations), we remark that the $z$-expansion fit to PRad data results in a significantly larger 
uncertainty for $r_E^p$ than is obtained using \eqnref{rational11}, comparable
to the $\sim 0.020\,{\rm fm}$ uncertainty of our default proton fit, while the combined
fit returns a radius uncertainty that is a factor of 2 smaller than either dataset in isolation.  

\section{Two-photon exchange} \label{app:TPE}

In this appendix, we provide pertinent details of two-photon exchange corrections 
that were discussed in Section~\ref{sec:radcorr}.

For momentum transfers $Q^2 \lesssim 1 \GeV^2$, dispersion relations have been used to 
constrain TPE corrections using available experimental data for inelastic cross sections. 
At relatively small momentum transfer and small scattering angles, the contribution from all inelastic intermediate states was evaluated~\cite{Tomalak:2015aoa} on top of the proton state~\cite{Blunden:2003sp} accounting for unpolarized proton inelastic structure functions in the resonance region. 
To calculate the TPE correction at large scattering angles, the data-driven dispersion relation framework was recently developed~\cite{Borisyuk:2008es,Tomalak:2014sva,Tomalak:2016vbf,Blunden:2017nby,Tomalak:2017shs}. 
The imaginary part of TPE amplitudes is evaluated from on-shell information in the physical region of electron--proton scattering.
The real part of TPE amplitudes requires information from the unphysical region as input. 
Novel methods of analytical continuation~\cite{Tomalak:2014sva,Tomalak:2017shs} allow us to overcome this complication. 
Contributions from proton and $\pi N$ intermediate states are evaluated for $Q^2 \lesssim 1 \GeV^2$.
At low momentum transfer and backward scattering angles, the relative contribution of inelastic intermediate states is found to be much smaller than the elastic contribution to TPE. 
At larger electron beam energies and momentum transfer, the intermediate states with higher invariant mass, e.g., $\pi \pi N$, become kinematically enhanced and prevent making a rigorous prediction in the absence of exclusive experimental data. 
At small momentum transfer $Q^2 \lesssim 0.25 \GeV^2$ and scattering angles, we account for all inelastic intermediate states~\cite{Tomalak:2015aoa}. 
At large angles and momentum transfer, proton and $\pi N$ states are included~\cite{Tomalak:2014sva,Tomalak:2016vbf,Tomalak:2017shs}.
The intermediate region is described by interpolation between these two calculations as in Ref.~\cite{Tomalak:2018ere}. We denote this dispersive result as $\delta^{\mathrm{dispersive}}$ and provide the corresponding correction for each point in the Mainz dataset in the Supplementary Material.

At larger momentum transfers, $Q^2 \gtrsim 1 \GeV^2$, the explicit form of the phenomenological 
TPE modification is as follows~\cite{arrington07c}:
\eqsl{arringtonTPE}{
\delta^{\mathrm{AMT}}(\varepsilon,~Q^2) &= 0.01 (\varepsilon - 1) \frac{\ln \frac{Q^2}{1 \GeV^2}}{\ln 2.2} \,,
}
which is negative (since $0 \leq \varepsilon \leq 1$) and increases the inferred Born cross section.
As discussed in the main text, 
this correction serves to improve agreement between polarization measurements and TPE-corrected 
unpolarized Rosenbluth measurements at high-$Q^2$.

\section{Comparison to literature} \label{app:complit}

In this appendix, we provide some existent results for form factor curvatures, Friar radii 
and Zemach radii.

\subsection{Curvature}

The curvature of the proton form factor has been estimated by performing fits to data~\cite{bernauer14:private} and by performing calculations in heavy-baryon ChPT~\cite{Horbatsch:2016ilr}. 
We tabulate these previous results in \tabref{curvatureslit}.
Our extraction of electric curvature lies below previous extractions from data. The curvature of neutron form factors was evaluated in dispersively-improved chiral effective field theory (DI$\chi$EFT). Our results for the curvatures of both electric and magnetic neutron form factors are in a fair agreement with Ref.~\cite{Alarcon:2018irp}.

\begin{table}[ht]
\centering
\caption{Curvature of proton $(p)$ and neutron $(n)$ electromagnetic form factors.
}
\label{tab:curvatureslit}
\begin{minipage}{\linewidth}
\centering
\begin{tabular}{|l|c|c|c|c|c|}
  \hline
  Fit choice & $\mean{r^4}^p_E$ $[\mathrm{fm^4}]$ & $\mean{r^4}^p_M$ $[\mathrm{fm^4}]$ & $\mean{r^4}^n_E$ $[\mathrm{fm^4}]$ & $\mean{r^4}^n_M$ $[\mathrm{fm^4}]$
\\
\hline
This paper & $1.08(28)(5)$ & $-2.0(1.7)(0.8)$ & $-0.33(24)(3)$ & $-2.3(2.1)(1.1)$ \\
Heavy-baryon ChPT~\cite{Horbatsch:2016ilr}&  $0.60(29)$ & $0.79(28)$ & --- & ---\\
DI$\chi$EFT~\cite{Alarcon:2018irp} &  $1.47 \cdots 1.60$ & $1.68 \cdots 1.78$ &$-0.64 \cdots -0.51$ & $2.04$\\
Fit to data~\cite{Higinbotham:2019jzd} &  $1.53$ & $0.91$ & --- & ---\\
Based on A1 fits~\cite{bernauer14:private} &  $2.31 \cdots 2.64$ & $-0.12 \cdots 0.75$ & --- & ---\\
\hline 
\end{tabular}
\end{minipage}
\end{table}

\subsection{Friar radius}

We present some previous estimates of Friar radii in Table~\ref{Friar_radii}.
There is a significant difference between results with and without the constraint on the proton charge radius~\cite{Friar:1997tr,Pachucki:1996zza,Pachucki:1999zza,Pineda:2004mx,Distler:2010zq,DeRujula:2010dp,Cloet:2010qa,Wu:2011jm,Indelicato:2012pfa,Borie:2012zz,Peset:2014jxa,Lorenz:2014yda,Karshenboim:2015bwa,Friar:2005jz,Graczyk:2015kka}.

\begin{table}[ht]
\centering
\caption{Friar radius of proton.}
\label{Friar_radii}
\begin{minipage}{\linewidth}  
\centering
\begin{tabular}{|l|l|l|}
  \hline          
   &$ \pr{r_F^p}^3~[\mathrm{fm}^3]$ 
\\
\hline
This paper &  $2.246(58)(2)$ \\
Using A1 fits~\cite{Distler:2010zq}&  $2.85(8)$ \\
Friar and Sick~\cite{Friar:2005jz}&  $2.71(13)$ \\
Graczyk and Juszczak~\cite{Graczyk:2015kka}&  $2.889(8)$ \\
\hline 
\end{tabular}
\end{minipage}
\end{table}

\subsection{Zemach radius}
    
Previous results for the nucleon Zemach radii are
$r^p_Z = 1.045(4)~\mathrm{fm}$ in Ref.~\cite{Distler:2010zq} for the proton and
$r^n_Z = -0.0449(13)~\mathrm{fm}$ in Refs.~\cite{Friar:2005je,Tomalak:2018ysg} for the neutron.
These should be compared with our values: $r^p_Z = 1.0227(94)(51)~\mathrm{fm}$ for the proton and $r^n_Z = -0.0445(14)(3)~\mathrm{fm}$ for the neutron.
Further calculations and fits to scattering data are found in Refs.~\cite{Pachucki:1996zza,Friar:2003zg,Friar:2005yv,Friar:2005je,Carlson:2008ke,Distler:2010zq,Graczyk:2015kka,Tomalak:2017npu}. 
Extractions from atomic spectroscopy are found in Refs.~\cite{antognini13,Dupays:2003zz,Volotka:2004zu,Hagelstein:2017cbl,Dorokhov:2017nzk}.

\bibliography{vector_ff_prd}

\end{document}


\maketitle

This document provides an explanation of the \texttt{.dat} files included in the Supplementary Material.

The file \texttt{Mainz\char`_ep\char`_CrossSections\char`_Rebinned\char`_v2.dat} is identical to that provided in the Supplemental Material of Phys.\ Lett.\ B {\bf 777}, 8 (2018), except that an additional column for the dispersive calculation of TPE corrections has been inserted as the third-last column from the right in the file, with heading \texttt{disp\char`_TPE\char`_corr}.

Below, we list the files that provide the fit values for coefficients corresponding to Eqs.~(13)--(16) in Sec.~2.5 of the paper with $k\ld{max} = 8$.
These are listed with 12 digits of precision after the decimal.
One major difference is the magnetic coefficients of the corresponding form factors are listed as
\eq{
G_M^N (z) = \sum_{k=0}^{k\ld{max}} \tilde{b}_k z(Q^2)^k \,,
}
i.e., the coefficients $\tilde{b}_k$ are multiplied by $G_M^N(0)$ relative to the $b_k$ as defined in the middle of Eq.~(4).
As explained in Secs.~2.1 and 2.4, we choose the 4 coefficients not determined by sum rules as $a_1, \ldots, a_4$, with $a_0$ and $a_5, \ldots, a_8$ determined by the normalization $G(0)$ and the sum rules in Eq.~(6).
\begin{enumerate}
\item \texttt{proton\char`_coefs\char`_kmax8.dat}: the first line lists $a_1^p, \ldots, a_4^p$, the second line lists $\tilde{b}_1^p, \ldots, \tilde{b}_4^p$ correspodning to Eq.~(13).
\item \texttt{GEn\char`_coefs\char`_kmax8.dat}: $a_1^n, \ldots, a_4^n$ in the top line of Eq.~(14).
\item \texttt{GMn\char`_coefs\char`_kmax8.dat}: $\tilde{b}_1^n, \ldots, \tilde{b}_4^n$ as in the bottom line of Eq.~(14).
\item \texttt{iso1\char`_coefs\char`_kmax8.dat}: the top two lines are the isoscalar coefficients $a_1^S, \ldots, a_4^S$ and $\tilde{b}_1^S, \ldots, \tilde{b}_4^S$ corresponding to Eq.~(16) and the bottom two lines are the isovector coefficients $a_1^V, \ldots, a_4^S$ and $\tilde{b}_1^V, \ldots, \tilde{b}_4^V$ corresponding to Eq.~(15).
\end{enumerate}
Likewise, the 5 free coefficients for the $k\ld{max} = 9$ case are listed in the corresponding \texttt{.dat} files as above with \texttt{kmax9.dat} replacing \texttt{kmax8.dat}.

The covariance matrices for the $k\ld{max} = 8$ fit are provided in the files as described below.
\begin{enumerate}
\item \texttt{proton\char`_cov.dat}: $8\times 8$ matrix for the ordering $a_1^p, \ldots, a_4^p, \tilde{b}_1^p, \ldots, \tilde{b}_4^p$.
\item \texttt{GEn\char`_cov.dat}: $4 \times 4$ matrix for $a_1^n, \ldots, a_4^n$.
\item \texttt{GMn\char`_cov.dat}: $4 \times 4$ matrix for $\tilde{b}_1^n, \ldots, \tilde{b}_4^n$.
\item \texttt{iso1\char`_cov.dat}: $16 \times 16$ matrix for $a_1^S, \ldots, a_4^S, \tilde{b}_1^S, \ldots, \tilde{b}_4^S, a_1^V, \ldots, a_4^S, \tilde{b}_1^V, \ldots, \tilde{b}_4^V$.
\end{enumerate}